\documentclass[12pt]{article}
\usepackage{amssymb,amsmath,amsthm}
\usepackage{amsfonts,bbm}
\usepackage{blindtext}
\usepackage{booktabs}
\usepackage{bm}
\usepackage{caption}
\usepackage{comment}
\usepackage{enumitem}
\usepackage{float}
\usepackage{graphicx}
\graphicspath{ {../figures/} }
\usepackage{indentfirst}
\usepackage{mathtools}
\usepackage{multirow}
\usepackage{placeins}
\usepackage{ragged2e}
\usepackage{siunitx}
\newcolumntype{d}{S[input-symbols = ()]}
\usepackage{xr}
\usepackage[bibstyle=numeric, citestyle=authoryear, doi=false, url=true, backend=biber, maxbibnames=10, maxcitenames=4, uniquelist=false, uniquename=false, sorting=nyt]{biblatex}
\renewbibmacro{in:}{}
\DeclareNameAlias{default}{family-given/given-family}
\newcommand{\citet}{\textcite}
\newcommand{\indicator}[1]{\mathbbm{1}\{#1\}}

    \def\independenT#1#2{\mathrel{\setbox0\hbox{$#1#2$}%
    \copy0\kern-\wd0\mkern4mu\box0}} 
\newcommand{\E}{\mathbb{E}}
\renewcommand{\P}{\mathrm{P}}

\usepackage{array}
\newcolumntype{L}[1]{>{\raggedright\arraybackslash}m{#1}}

\usepackage{rotating}
\usepackage{setspace}
\usepackage{setspace}
\usepackage{subcaption}
\usepackage{ctable}
\usepackage[colorlinks,citecolor=blue,urlcolor=blue,hyperfootnotes=false]{hyperref}
\usepackage{cleveref}
\usepackage{dcolumn}
\usepackage{float}
\newcommand{\e}[1]{{\mathbb E}\left[ #1 \right]}
\usepackage[utf8]{inputenc}
\usepackage[margin=.7in]{geometry}
\usepackage{thmtools}

\makeatletter

\newcommand{\Rmnum}[1]{\expandafter\@slowromancap\romannumeral #1@}
\makeatother

\newtheorem{theorem}{Theorem}

\newtheorem{assumption}{Assumption}

\newtheorem{remark}{Remark}
\newtheorem{proposition}{Proposition}

\newtheorem{example}{Example}

\newtheorem{inneruassumption}{Assumption}
\newenvironment{uassumption}[1]
  {\renewcommand\theinneruassumption{#1}\inneruassumption}
  {\endinneruassumption}

\crefname{figure}{Figure}{Figures}
\crefname{assumption}{Assumption}{Assumptions}
\crefname{inneruassumption}{Assumption}{Assumptions}

\numberwithin{sassumption}{section}
\numberwithin{stheorem}{section}

\newfloat{sfigure}{tbp}{fig}

\begin{document}
\title{Treatment Effects in Interactive Fixed Effects Models with a Small Number of Time Periods\thanks{We thank the editor and two anonymous referees as well as Nicholas Brown, Weige Huang, Pedro Sant'Anna, Emmanuel Tsyawo, Jeffrey Wooldridge, and seminar participants at Michigan State University, the University of S\~{a}o Paulo, the 26th International Conference on Panel Data, and the 2020 Southern Economics Association Conference for helpful comments.  A previous version of this paper was circulated with the title ``Treatment Effects in Interactive Fixed Effects Models.'' Code for the approach suggested in the current paper is available in the \texttt{R} package \texttt{ife} which is available at \url{https://github.com/bcallaway11/ife}.}}
\author{Brantly Callaway\thanks{Assistant Professor, Department of Economics, University of Georgia, Email: brantly.callaway@uga.edu.} \and Sonia Karami\thanks{Senior Quantitative Analyst / Data Scientist, Quantitative and Supervision Research (QSR) Department, Federal Reserve Bank of Richmond, Email: sonia.karami@rich.frb.org.}}
\date{\today}

\maketitle

\abstract{\noindent This paper considers identifying and estimating the Average Treatment Effect on the Treated (ATT) when untreated potential outcomes are generated by an interactive fixed effects model.  That is, in addition to time-period and individual fixed effects, we consider the case where there is an unobserved time invariant variable whose effect on untreated potential outcomes may change over time and which can therefore cause outcomes (in the absence of participating in the treatment) to follow different paths for the treated group relative to the untreated group.  The models that we consider in this paper generalize many commonly used models in the treatment effects literature including difference in differences and individual-specific linear trend models.  Unlike the majority of the literature on interactive fixed effects models, we do not require the number of time periods to go to infinity to consistently estimate the ATT.  Our main identification result relies on having the effect of some time invariant covariate (e.g., race or sex) not vary over time.  Using our approach, we show that the ATT can be identified with as few as three time periods and with panel or repeated cross sections data.}

\vspace{30pt}


\bigskip


\vspace{120pt}

\pagebreak

\normalsize

\onehalfspacing

\section{Introduction}

One of the most common ways to identify the causal effect of a binary treatment (e.g., participating in a program or being affected by some economic policy) on an outcome of interest is to exploit variation in the timing of treatment across different individuals.  In this sort of setup, the central idea is to impute the average outcome that individuals that participate in the treatment \textit{would have experienced} if they had not participated in the treatment using a combination of their observed pre-treatment outcomes and the path of outcomes for a group of individuals that does not participate in the treatment.  One can then estimate the average effect of participating in the treatment (for the group of individuals that participate in the treatment) as the difference in average observed outcomes and the average imputed value for individuals that participate in the treatment.

The most common version of this approach is difference in differences (DID) where the key identifying assumption is that, in the absence of treatment, the ``path'' of outcomes that the treated group would have experienced is the same, on average, as the path of outcomes that an untreated group did experience.
The main motivating model for the DID approach is one in which untreated ``potential'' outcomes are generated by a two-way fixed effects model; that is, a model that allows for time invariant unobserved heterogeneity that may be distributed differently across the treated and untreated groups as well as allowing for time fixed effects.  This model is consistent with the so-called ``parallel trends'' assumption underlying the DID approach (see, for example, \citet{blundell-dias-2009,borusyak-jaravel-spiess-2021,gardner-2021}).  However, the parallel trends assumption may not hold in many applications; often, a main concern is that the effect of some unobserved time invariant variable may change over time which would cause the parallel trends assumption to be violated.  In those cases, researchers often resort to other approaches such as including individual-specific linear trends (\citet{heckman-hotz-1989,wooldridge-2005,mora-reggio-2019}).

We generalize both the two-way fixed effects model and these alternative approaches by imposing that the model for untreated potential outcomes has an interactive fixed effects structure (also known as factor structure).  These models allow for the effect of time invariant unobservables to change over time.  In particular, we consider the following sort of model for untreated potential outcomes:
\begin{align} \label{eqn:1}
  Y_{it}(0) = \theta_t + \xi_i + \lambda_i F_t + U_{it}  
\end{align}
where $Y_{it}(0)$ is an individual's untreated potential outcome in time period $t$, $\theta_t$ is a time fixed effect, $(\xi_i,\lambda_i)$ are unobserved, time invariant individual characteristics, and $F_t$ is a time varying effect of $\lambda_i$.  Notice that this model generalizes the common two-way fixed effects model; in particular, they are the same when $F_t$ does not change across time (implying that the term $\xi_i + \lambda_i F_t$ does not change over time and standard DID approaches can be applied).  This model also generalizes individual-specific linear trend models that are commonly used in applied work; in particular, this would be true when $F_t = t$ in all time periods.\footnote{The model in \Cref{eqn:1} includes a single interactive fixed effect.  This is a leading case as it generalizes all the most common approaches taken in the treatment effects literature; also, it corresponds to more structural interpretations where the researcher is concerned about a particular unobserved time invariant variable whose effect can change over time.  In \Cref{sec:multiple-ife}, we discuss the case when there can be more than one interactive fixed effect.}  A main motivation for this sort of model is one where a researcher worries that (i) there are time invariant unobserved variables that affect the outcome of interest that are not observed in the data, and (ii) the effect of one of those time invariant unobservables varies over time.

In the current paper, we develop a new approach to identifying treatment effects when untreated potential outcomes are generated from an interactive fixed effects model as in \Cref{eqn:1}.    
Unlike previous work on treatment effects in interactive fixed effects models, our approach does not require the number of time periods to go to infinity to consistently estimate the effect of participating in the treatment.  We consider the case where a researcher observes a vector of time invariant covariates (e.g., race, sex, or education).  The key assumption of our main approach is that the effect of at least one of the covariates on untreated potential outcomes does not change over time.  The effect of this covariate is not identified, but we show that it can still be useful for identifying the effect of participating in the treatment.  This sort of covariate is likely to be available in many applications in economics as it would be available in any model for untreated potential outcomes where some particular covariate ``drops out'' due to not varying over time (up to an instrument relevance condition that can be checked with the available data).  In this case, we show that the ATT is identified with only three periods (this is the same requirement as for individual-specific linear trend models and one more time period than is required for DID models).  In practice, our approach amounts to using this time invariant covariate as an instrument for the change in outcomes over time in a regression that ``differences out'' the individual fixed effects.  This time invariant covariate is allowed to affect the level of untreated potential outcomes (just not to directly affect the path of untreated potential outcomes over time); we also do not use it as an instrument for participating in the treatment.  Interestingly, this approach can be implemented with repeated cross sections data.  We appear to be the first paper to propose a method allowing for interactive fixed effects that can be implemented with only repeated cross sections data.

We also develop estimators of the ATT using our approach.  We propose a two-step estimator where, in the first step, we estimate the parameters in the interactive fixed effects model for untreated potential outcomes.  In the second step, we plug these in to obtain estimates of the ATT.  In the case with exactly three time periods and one covariate whose effect does not change over time, the ATT is exactly identified.  With more than one covariate whose effect does not change over time, the ATT is over-identified and the restrictions that the effects of covariates do not vary over time can be tested.  With more than three periods, one can conduct ``pre-tests'' for (i) the effect of participating in the treatment being 0 in periods before individuals become treated as well as for (ii) the effect of particular covariates not varying over time.  We also extend our baseline results to the case where there is variation in treatment timing across individuals (as has been considered in recent work on DID such as \citet{goodman-2021,callaway-santanna-2021}).

We conclude the paper with an application on the effect of early-career job displacement on earnings.  In the job displacement literature, researchers are typically concerned that ``skill'' is unobserved and cannot be directly controlled for.  If the return to skill varies over time or with experience, then standard approaches such as difference in differences are likely to lead to biased estimates of the effect of job displacement.  On the other hand, the approach proposed in the current paper can deal directly with this issue.  Compared to difference in differences, the interactive fixed effects approach proposed in the current paper does somewhat better in pre-treatment periods (in the sense of estimating no effect of job displacement in pre-displacement periods) while finding somewhat smaller long-term effects of job displacement.

The outline of the paper is as follows.  Section 2 provides our main identification arguments.  Section 3 discusses some limitations of difference in differences and linear trends approaches in the context of interactive fixed effects models for untreated potential outcomes.  Section 4 proposes an estimation and inference strategy.  Section 5 provides a number of extensions to our main identification strategy.  Section 6 provides Monte Carlo simulations related to comparing our approach to difference in differences and linear trends.  Section 7 includes an application on job displacement.  The Supplementary Appendix provides some alternative identification strategies; additional Monte Carlo simulations concerning (i) weak instruments, (ii) estimation with multiple interactive fixed effects, and (iii) selecting the number of interactive fixed effects; and some additional results for our application on job displacement. 

\paragraph{Related Literature} \ \mbox{}

There is a large literature on linear models with interactive fixed effects.  Important early work in this literature includes \citet{pesaran-2006,bai-2009}.  For estimation, these papers and much subsequent work require a large number of time periods (i.e., the number of time periods goes to infinity).  Our paper is particularly related to \citet{gobillon-magnac-2016,xu-2017}.  Like the current paper, these papers model untreated potential outcomes by imposing an interactive fixed effects structure.  In the case where a researcher is interested in the effect of a binary treatment, this approach has a major advantage over imposing linear models for all outcomes --- this setup places no restrictions on how treated potential outcomes are generated at all.  Thus, one can allow for unrestricted forms of treatment effect heterogeneity, very general forms of selection into treatment, and treatment effect dynamics.   The key difference between our approaches are that we do not require a large number of time periods, but we do require access to a covariate whose effect does not change over time.  Our approach should ultimately be seen as complementary to the approaches taken in those papers and, in particular applications,  the relative merits of these approaches mainly comes down to data availability and the plausibility of the condition that effects of covariate(s) do not vary over time.

The current paper is also closely related to work on factor models with a small number of measurements (for example, \citet{anderson-rubin-1956,heckman-scheinkman-1987,carneiro-hansen-heckman-2003,heckman-stixrud-urzua-2006,williams-2020}).  Related work on panel data models with interactive fixed effects and with a small number of periods includes \citet{holtz-newey-rosen-1988,ahn-lee-schmidt-2001,ahn-lee-schmidt-2013,robertson-sarafidis-2015,freyberger-2018,juodis-sarafidis-2020}.  Of these, the most similar is \citet{ahn-lee-schmidt-2013}; like that paper, we obtain identification using a ``differencing'' approach and by providing some extra moment conditions.  Our paper is different in that we specifically focus on a treatment effects setup (e.g., our approach allows for much more general forms of treatment effect heterogeneity than simply including a treatment dummy variable directly in an interactive fixed effects model).  We also focus on the case with time invariant covariates (discussed in more detail below) rather than time varying covariates.  Our paper is also related to \citet{freyaldenhoven-hansen-shapiro-2019}.  That paper deals with a more general form of unobservables (time-varying rather than interactive fixed effects) that confound identifying the effect of the treatment; on the other hand, our approach allows for the excluded variable to be time invariant and to directly affect the outcome as well as allowing for more general treatment effect heterogeneity (though it would seem possible to extend their approach along this dimension).  \citet{gardner-2020} allows for violations of parallel trends when individuals belong to a latent class in the case where there are a small number of time periods.  Finally, a number of other approaches (particularly in the literature on synthetic controls and matrix completion) are motivated by an underlying interactive fixed effects model; see, for example, the discussion in \citet{abadie-diamond-hainmueller-2010}.  These include \citet{hsiao-ching-wan-2012,kim-oka-2014,li-bell-2017,hsiao-zhou-2019,powell-2019,arkhangelsky-etal-2021,ben-feller-rothstein-2021,ferman-pinto-2021,kellogg-mogstad-pouliot-torgovitsky-2021,athey-bayati-doudchenko-imbens-khosravi-2021,bai-ng-2021,liu-wang-xu-2021}, among others.

\section{Identification}

\subsection{Notation and Parameters of Interest}

We use potential outcomes notation throughout the paper.  Let $Y_{t}$ denote an individual's outcome in a particular period $t$ (we drop an individual subscript $i$ throughout much of the paper except where it enhances clarity).  We assume that a researcher has access to $\mathcal{T}$ periods of data.  Individuals either belong to a treated group or untreated group.  We set $D$ to be a treatment indicator so that $D=1$ for individuals in the treated group and $D=0$ for the untreated group.  We assume that treatment occurs in period $t^*$ which is the same across all individuals (we discuss the case with variation in treatment timing in \Cref{sec:staggered}).  Individuals have treated potential outcomes (the outcomes that would occur if an individual were treated) and untreated potential outcomes (the outcomes that would occur if an individual were not treated) at each time period.  We denote these by $Y_t(1)$ and $Y_t(0)$, respectively, for $t=1,\ldots,\mathcal{T}$.  In pre-treatment periods, i.e. $t < t^*$, $Y_t = Y_t(0)$ for all individuals; that is, we observe untreated potential outcomes for all individuals in these periods.  When $t \geq t^*$, $Y_t = D Y_t(1) + (1-D)Y_t(0)$; that is, in period $t^*$ and subsequent time periods, we observe treated potential outcomes for individuals that participate in the treatment and untreated potential outcomes for individuals that do not participate in the treatment.

We also suppose that we observe a $K \times 1$ vector of covariates $Z$ that are time invariant.  Time invariance of the covariates is essentially standard in the literature on treatment effects with panel data (\citet{heckman-ichimura-smith-todd-1998,abadie-2005,callaway-santanna-2021}).\footnote{There is an interesting difference between the econometrics literature and common practices in applied work in the presence of time varying covariates.  The econometrics literature typically conditions on a single value of pre-treatment covariates in this case.  This approach can allow for the treatment to have an effect on the covariates themselves; moreover, it can be rationalized under additional assumptions on how time-varying covariates would have evolved in the absence of participating in the treatment.  On the other hand, applied work most often estimates two way fixed effects regressions that only include covariates that vary over time and that rely on additional functional form assumptions as well as exogeneity assumptions that rule out the treatment having an effect on the covariates.}  Unlike traditional panel data models, the objective in the treatment effects literature is not to estimate the effect of the covariates, but simply to control for them.  Thus, time invariant (or pre-treatment) covariates allow us to compare outcomes or the path of outcomes for individuals who have the same characteristics rather than focusing on using variation in these covariates over time to identify their effects.  A key difference between our approach and standard two-way fixed effects regressions is that we treat the treatment variable asymmetrically from the other covariates.  Another concern about time varying covariates is that participating in the treatment may affect the path of some time varying covariates (\citet{lechner-2008,bonhomme-sauder-2011}).  It seems possible to extend our results to allow for time varying covariates along the lines of marginal structural models (\citet{robins-hernan-brumback-2000,imai-ratkovic-2015,blackwell-glynn-2018}), but we do not pursue this here. 

The main parameter that we are interested in is the Average Treatment Effect on the Treated (ATT); we define $ATT_t$ as the average treatment effect on the treated in period $t$.  That is,
\begin{align*}
  ATT_t = \E[Y_t(1) - Y_t(0)|D=1]
\end{align*}
$ATT_t$ allows researchers to consider how treatment effects vary across time periods / length of exposure to the treatment.  When $ATT_t$ is identified, it can also be aggregated into an overall $ATT$ across all post-treatment time periods
\begin{align*}
    ATT = \frac{1}{\mathcal{T}-t^*+1} \sum_{t=t^*}^{\mathcal{T}} ATT_t
\end{align*}
which is just the average of $ATT_t$ across all post-treatment time periods.

The first main assumption is about the data generating process for untreated potential outcomes.  In particular, we impose the following structure on untreated potential outcomes in each time period.
\begin{assumption}[Model for Untreated Potential Outcomes]\label{ass:model}  Untreated potential outcomes are generated by the following interactive fixed effects model
\begin{align*} 
  Y_{it}(0) = \xi_i + \lambda_i F_t + Z_i'\delta_t + U_{it}  
\end{align*}
\end{assumption}
We assume that in all time periods, untreated potential outcomes are generated by an interactive fixed effects model that includes an individual fixed effect, a time fixed effect (to conserve on notation we absorb the time fixed effect into the term $Z_i'\delta_t$ by having $Z_i$ include an intercept and where the corresponding element of $\delta_t$ is equal to the time fixed effect), and a single time invariant unobservable whose ``effect'' can change over time.  

Some important comments about the model in \Cref{ass:model} are in order.  First, \Cref{ass:model} only puts structure on how untreated potential outcomes are generated.  It does not put any structure on how treated potential outcomes are generated, and this allows for essentially unrestricted treatment effect heterogeneity.  In particular, our setup allows for individuals to select into the treatment on the basis of having ``good'' treated potential outcomes relative to individuals who do not participate in the treatment.  Not putting any structure on treated potential outcomes comes at the cost of not identifying more general parameters such as the average treatment effect (for the entire population).  In particular, this means that it may be difficult to predict what the effect of participating in the treatment would have been for individuals that did not participate in the treatment; however, this is a standard issue for DID-type methods.  Second, \Cref{ass:model} allows for the distributions of $\xi$, $\lambda$, and $Z$ to be different for individuals in the treated group relative to the untreated group.  In the spirit of fixed effects models, \Cref{ass:model} also allows for these variables to be arbitrarily correlated with each other.  

Third, having the coefficient on $Z$ vary over time follows the difference in differences literature.  If one removes the interactive fixed effect term from the model in \Cref{ass:model}, then this is exactly the sort of model that gives rise to conditional parallel trends assumptions that are common in the econometrics literature (\citet{heckman-ichimura-smith-todd-1998,abadie-2005,callaway-santanna-2021}).  Including the interactive fixed effect puts the unobserved $\lambda$ into the model in the same way as the covariates.  This means that, if we were to observe both $\lambda$ and $Z$, then the path (not the level) of untreated potential outcomes that individuals in the treated group would have experienced if they had not not participated in the treatment is the same as the path of outcomes that individuals in the untreated group did experience \textit{conditional on having the same values of $\lambda$ and $Z$}.  In other words, if we observed $\lambda$, then we could just use a conditional difference in differences approach, but, since we do not observe it, we need to make some modifications.  Finally, \Cref{ass:model} restricts the dimensionality of the interactive fixed effect term to be equal to one.  We discuss the extension to multiple interactive fixed effects in \Cref{sec:multiple-ife}.

We also make the following assumptions throughout the paper.

\begin{assumption}[Observed Data]  \label{ass:sampling} The observed data consists of iid draws of $\{Y_{i1},Y_{i2},\ldots,Y_{i\mathcal{T}}, Z_i, D_i\}_{i=1}^n$, where $n$ denotes the sample size.  
\end{assumption}

\begin{assumption}[Treatment Timing] \label{ass:timing} $\mathcal{T} \geq 3$ and $t^* \geq 3$. 
\end{assumption}

\begin{assumption}[Factors] \label{ass:factors} $F_{t^*-1} \neq F_{t^*-2}$.
\end{assumption}

\begin{assumption}[Selection on observables and time invariant unobservables] \label{ass:selection}
  For all $t=1,\ldots,\mathcal{T}$,
  \begin{align*}
    \E[Y_t(0)|\xi,\lambda,Z,D=1] = \E[Y_t(0)|\xi,\lambda,Z,D=0]
  \end{align*}
\end{assumption}
\Cref{ass:sampling} says that we observe panel data with $\mathcal{T}$ time periods.  We discuss extending the results to the case with repeated cross sections in \Cref{sec:repeated-cross-sections} and note here that this extension is straightforward. \Cref{ass:timing} says that we have access to at least three periods of data, that treatment starts in period $t^*$ for all individuals that participate in the treatment, and that we have at least two pre-treatment periods for individuals in the treated group.  \Cref{ass:factors} says that there is variation in $F_t$ in the two periods immediately preceding treatment.  Practically, this assumption implies that differences in $\lambda$ in pre-treatment periods lead to differences in average paths of outcomes in pre-treatment periods.  \Cref{ass:selection} is the same as Assumption 2 in \citet{gobillon-magnac-2016} and is similar to Assumption 1 in \citet{gardner-2020}.  Moreover, it is similar to the assumption of selection on observables that is commonly invoked in the treatment effects literature with cross sectional data (see, for example, \citet{imbens-wooldridge-2009}); however, in our case, \Cref{ass:selection} holds after conditioning on the unobserved heterogeneity including the time invariant unobservable whose effect can change across time.  In particular, it says that our main identification challenge is due to $\xi$ and $\lambda$ being unobserved.  It implies that, if one could observe and condition on both $\xi$ and $\lambda$, that average untreated potential outcomes would be the same in the treated and untreated groups.  This would also imply that, were the unobserved heterogeneity and interactive fixed effect observed, one could use a variety of well known approaches (e.g., least squares regression, propensity score re-weighting, or matching) to estimate the ATT.  In our case, where these time invariant variables are unobserved, this assumption is potentially helpful because untreated potential outcomes are not observed for the treated group (in some periods), but they are observed for the untreated group in all periods.

In the context of the model in \Cref{ass:model}, an important implication of \Cref{ass:selection} is that it is equivalent to
\begin{align}\label{eqn:1a}
  \E[U_{t}|\xi,\lambda,Z,D=d] = 0 \quad \text{for all} \ (t,d) \in \{1,2,\ldots,\mathcal{T}\} \times \{0,1\}
\end{align}
\Cref{eqn:1a} immediately implies that the time varying unobservables in \Cref{ass:model} are uncorrelated with the observed covariates, with the time invariant unobservables in the model, and with whether or not an individual participates in the treatment.  These are all typical assumptions in the interactive fixed effects literature.  That these unobservables are uncorrelated with covariates and treatment status will eventually be a source of moment conditions that we use to identify the parameters in the model.  \Cref{eqn:1a} also makes clear the main type of restriction that we place on how individuals select into participating in the treatment.  In addition to allowing for individuals to select into the treatment on the basis of their treated potential outcomes, we allow for individuals to select into treatment on the basis of their time invariant unobservables that affect untreated potential outcomes $(\xi,\lambda)$, but we do not allow individuals to select into treatment on the basis of time varying unobservables, $U_t$.  

Next, we outline the challenges that a researcher faces for identifying $ATT_t$ when untreated potential outcomes are generated by the interactive fixed effects model in \Cref{ass:model}.  As a first step, we discuss a  ``quasi-differencing'' strategy in order to write down equations that do not depend on $\xi$ or $\lambda$.\footnote{Quasi-differencing is common in the interactive fixed effects literature especially in the case with a ``small'' number of time periods (see, for example, \citet{holtz-newey-rosen-1988,ahn-lee-schmidt-2013}).}  Notice that, given the model in \Cref{ass:model}, 
\begin{align} \label{eqn:dy}
  Y_{it^*-1}(0) - Y_{it^*-2}(0) &= \lambda_i \big(F_{t^*-1} - F_{t^*-2}\big) + Z_i' \big(\delta_{t^*-1} - \delta_{t^*-2} \big) + U_{it^*-1} - U_{it^*-2}
\end{align}
Because no one is treated before period $t^*$, the researcher observes untreated potential outcomes in the two time periods immediately preceding treatment for both treated and untreated individuals.  Further, one can immediately solve this equation for $\lambda_i$:
\begin{align} \label{eqn:lam}
  \lambda_i &= \Big( \big( Y_{it^*-1}(0) - Y_{it^*-2}(0) \big)  - Z_i' \big(\delta_{t^*-1} - \delta_{t^*-2}\big) - \big(U_{it^*-1}-U_{it^*-2}\big) \Big) \Big/ (F_{t^*-1} - F_{t^*-2})
\end{align}
which follows by re-arranging the expression in \Cref{eqn:dy}.  Using similar arguments, it additionally follows that 
\begin{align} \label{eqn:2}
    Y_{it}(0) - Y_{it^*-2}(0) &= \lambda_i(F_t - F_{t^*-2}) + Z_i'(\delta_t - \delta_{t^*-2}) + U_{it} - U_{it^*-2} \nonumber \\
    &= Z_i'\delta^*_t + F^*_t (Y_{it^*-1} - Y_{it^*-2}) + V_{it}
\end{align}
where we define $F^*_t := \big(F_t-F_{t^*-2}\big) \Big/ \big(F_{t^*-1} - F_{t^*-2}\big)$, $\delta^*_t := \big(\delta_t - \delta_{t^*-2} - F^*_t(\delta_{t^*-1} - \delta_{t^*-2}) \big)$, $V_{it} := (U_{it} - U_{it^*-2}) - F^*_t(U_{it^*-1} - U_{it^*-2})$ and where the second equality holds by plugging in the expression for $\lambda_i$ in \Cref{eqn:lam} and re-arranging terms.  The result in \Cref{eqn:2} holds for all $t = t^*,\ldots,\mathcal{T}$ for both the treated group and the untreated group, and, importantly, \Cref{eqn:2} has differenced out the time invariant unobservables. 

Next, in order to think about identifying $ATT_t$, notice that, for any time period $t \geq t^*$,
\begin{align} \label{eqn:att-id-challenge}
    ATT_t &= \E[Y_t(1) - Y_t(0) | D=1] \nonumber \\
    &= \E[Y_t(1) - Y_{t^*-2}(0) | D=1] - \E[Y_t(0) - Y_{t^*-2}(0) | D=1] \nonumber \\
    &= \E[Y_t - Y_{t^*-2} | D=1] - \Big( \E[Z'|D=1] \delta^*_t + F^*_t \E[Y_{t^*-1} - Y_{t^*-2} | D=1] \Big)
\end{align}
where the second equality holds by adding and subtracting $\E[Y_{t^*-2}(0) | D=1]$, and the third equality holds by writing potential outcomes in terms of their observed counterparts, by \Cref{eqn:2}, and by Assumptions \ref{ass:model} and \ref{ass:selection}.  \Cref{eqn:att-id-challenge} suggests that the key identification challenge is to be able to recover the parameters $\delta^*_t$ and $F^*_t$; all the other terms in \Cref{eqn:att-id-challenge} are immediately identified from the sampling process.  It is also worth pointing out that identifying $F^*_t$ is not as challenging as identifying $F_t$ itself in all time periods, and, in particular, notice that we do not need to impose any of the normalizations on $F_t$ that are common in the interactive fixed effects literature.

Because, in the post-treatment periods $t=t^*,\ldots,\mathcal{T}$, $Y_{it}(0)$ is only an observed outcome for individuals in the untreated group, \Cref{eqn:2} therefore suggests recovering these parameters using the untreated group.  That being said, $\delta^*_t$ and $F^*_t$ are not immediately identified in \Cref{eqn:2} because $(Y_{it^*-1} - Y_{it^*-2})$ is correlated with $V_{it}$ by construction.  We discuss our approach to identifying $\delta^*_t$ and $F^*_t$ (and hence $ATT_t$) next.

\begin{remark}
  The above arguments relied on differencing both $Y_{it}(0)$ and $Y_{it^*-1}(0)$ by $Y_{it^*-2}(0)$, but there alternative differencing strategies that could eliminate the time invariant unobserved heterogeneity terms.  One advantage of the present strategy is that it only relies on the interactive fixed effects model for untreated potential outcomes holding from period $t^*-2$ to period $t$ and is therefore robust to violations of this model in earlier time periods.\footnote{Imposing that the structure of the model holds over a large number of time periods may be restrictive in many applications.  For example, in a related context, \citet{arkhangelsky-etal-2021} note that ``it may be difficult to find a simple specification...that fits well over the entire panel.''}  A main alternative would be to difference by $\bar{Y}^{pre(t^*-2)}_i := \frac{1}{t^*-2} \sum_{s=1}^{t^*-2} Y_{is}(0)$.  This strategy would require that the interactive fixed effects structure hold in all pre-treatment periods, though it would not require \Cref{ass:factors} (that $F_{t^*-1} \neq F_{t^*-2}$) but would instead require $F_{t^*-1} \neq \bar{F}^{pre}_{t^*-2} := \frac{1}{t^*-2} \sum_{s=1}^{t^*-2} F_s$; this might be attractive in cases where $F_t$ does not change across some time periods.
\end{remark}

\subsection{Identification through Covariates with Time Invariant Effects} \label{sec:2.3}

In this section, we present our main identification strategy.  As a first step, notice that \Cref{eqn:1a} combined with \Cref{eqn:2} provides a number of moment conditions that are relevant for identifying $(\delta^*_t,F^*_t)$.  In particular, the available moment conditions are
\begin{align*}
  0 &= \E[Z V_t | D=0] \ \textrm{for } t=t^*,\ldots,\mathcal{T} 
\end{align*}
These moment conditions are equivalent to
\begin{align} 
  0 &= \E[(1-D) Z ( Y_t - Y_{t^*-2} - Z'\delta^*_t - F^*_t(Y_{t^*-1} - Y_{t^*-2}))] \quad \textrm{for } t=t^*,\ldots,\mathcal{T} \label{eqn:momcond1}
\end{align}
To begin with (and for simplicity), consider the case where $\mathcal{T}=3$ (i.e., the researcher has access to three periods of panel data).  Then, \Cref{eqn:momcond1} contains $K$ moment conditions.  However, there are $K+1$ parameters to identify; $\delta^*_3$ is $K \times 1$ and $F^*_3$ is an extra scalar parameter to identify.  Hence, there are not enough moment equations to identify all the parameters.

Next, we introduce our main identifying assumption.

\begin{assumption}[Availability of Covariate(s) with Time Invariant Effects] \label{ass:special-cov} There exists a set $\mathcal{J} \subseteq \{1,\ldots,K\}$ such that $\delta_{jt} = \delta_j$ for all $j \in \mathcal{J}$ and for all time periods $t=1,\ldots,\mathcal{T}$. 
\end{assumption}

\Cref{ass:special-cov} introduces the central idea of our identification strategy.  In particular, we suppose that at least one of the $\delta_t$ parameters does not depend on time.  In many cases, this is a very weak requirement.  In particular, researchers often omit time invariant covariates because they are ``absorbed'' into the fixed effect --- this indeed does hold if the effect of some time invariant covariate does not vary over time.  Thus, any covariate that a researcher would otherwise omit from the model because it does not vary over time is a good candidate here.

To formalize things, set $Z=(X,W)$ which partitions the observed covariates into a group of $K_X$ covariates with time varying effects on the outcome and another group of $K_W$ covariates whose effects do not change over time; these satisfy $K=K_X + K_W$.  Correspondingly, partition $\delta_t = (\beta_t, \alpha)$ where $\beta_t$ is a $K_X$ dimensional vector and $\alpha$ is a $K_W$ dimensional vector that does not change across time periods, and likewise, partition $\delta^*_t = (\beta^*_t, 0)$.\footnote{\label{fn:delta_star_equals_0} Given the definition of $\delta_t^*$, if $\delta_{jt} = \delta_j$ for all time periods, then that immediately implies $\delta_{jt}^*=0$ for all time periods $t=t^*,\ldots,\mathcal{T}$.}  Then, we write the model in \Cref{ass:model} as
\begin{align} \label{eqn:3}
  Y_{it}(0) = \xi_i + \lambda_i F_t + X_i'\beta_t + W_i'\alpha + U_{it}
\end{align}
$\alpha$ is not identified in \Cref{eqn:3} and, in applications, information on $W$ is often completely discarded.  However, the presence of $W$ still provides potentially useful moment conditions.  In \Cref{eqn:3} (and in each time period), there are now $K_X + 1$ parameters to identify, and $K$ moment conditions available to identify them.  Then, as long as $K_W \geq 1$ (i.e., as long as there is at least one variable whose effect does not change over time), our approach satisfies the necessary order condition that $K \geq K_X+1$.

The simplest version of this approach is the case where we have exactly three time periods, panel data, and $W_i$ is a scalar.  Then, our idea amounts to estimating $\beta^*_3$ and $F^*_3$ using the sample of untreated individuals by estimating the model
\begin{align*}
  Y_{i3} - Y_{i1} = X_i'\beta^*_3 + F^*_3(Y_{i2}-Y_{i1}) + V_{i3}
\end{align*}
using $W_i$ as an instrument for $(Y_{i2} - Y_{i1})$.  In order for $\beta^*_3$ and $F^*_3$ to be identified here, we just require standard exogeneity and relevance conditions for instrumental variables estimation.  Exogeneity holds by \Cref{ass:special-cov} (along with \Cref{ass:selection}).  Relevance will hold if $\textrm{rank}(\E[ Z (X', (Y_{2} - Y_{1}))])=K_X +1$.  The main requirement here is that $W$ is correlated with $(Y_{2}- Y_{1})$ after controlling for $X$.  This condition can be checked in particular applications.\footnote{A relevant intermediate case is when $W$ is a weak instrument for $(Y_{2}-Y_{1})$.  Existing strategies for dealing with weak instruments can be applied in our setup.  For example, standard diagnostics for weak instruments (e.g., \citet{stock-yogo-2005,kleibergen-paap-2006,olea-pflueger-2013,sanderson-windmeijer-2016}) can be used.  Alternatively, one could employ estimation strategies designed for the case with weak instruments (e.g., \citet{hansen-kozbur-2014,andrews-armstrong-2017,tsyawo-2021}).}    In fact, an alternative equivalent condition is that $\textrm{rank}(\E[ Z (X', \lambda) ] = K_X+1$; this condition says that $W$ is correlated with $\lambda$ after controlling for $X$.  In other words, we need to find a variable whose effect on untreated potential outcomes does not change over time (i.e., this variable does not affect the path of untreated potential outcomes over time) and that is correlated with the time invariant unobservable whose effect changes over time.

To give an example, in the application in the paper, we study the effect of job displacement on earnings.  There, we are concerned that the return to ``skill'', which is unobserved, could vary over time.  In this case, there are several possibilities for which variables to impose that their effect on untreated potential outcomes does not change over time.  First, covariates such as race, sex, and education often ``drop out'' in applications on job displacement because they do not change over time.  This suggests that any of these are potential candidates here.\footnote{It's worth pointing out that just because a covariate does not vary over time does not automatically imply that the path of untreated potential outcomes does not depend on it.  In \Cref{sec:time-invariant-effects-test}, when there are extra available pre-treatment periods, we develop a test for covariates having time invariant effects in pre-treatment periods.  This test can be used as a check on the validity of the restriction that a particular covariate does not affect the path of untreated potential outcomes (given that $\lambda$ is included in the model).}  Another possibility is to use a proxy for unobserved skill.  In our application, we observe each individual's score on the Armed Forces Qualification Test (AFQT).  This variable can be seen as a proxy for unobserved skill, and it seems plausible that earnings in the absence of job displacement do not depend at all (notice that this is a stronger condition than our approach actually requires) on AFQT score once ``skill'' is in the model.

We state a general identification result for the $ATT_t$ next. Define
\begin{align*}
  \Delta Y^{post,t^*-2} := (Y_{t^*} - Y_{t^*-2}, \ldots, Y_{\mathcal{T}} - Y_{t^*-2})'
\end{align*} 
In other words, $\Delta Y^{post,t^*-2}$ is the difference between outcomes in particular post-treatment periods relative to the outcome two periods before the treatment started.  $\Delta Y^{post,t^*-2}$ is a $q \times 1$ vector where $q := (\mathcal{T}-t^*+1)$ which is the total number of post-treatment time periods.  Next, define
\begin{align*}
    \mathbf{Z} = \big( \mathbf{I}_{q} \otimes (1-D)Z \big)
\end{align*}
where $\mathbf{I}_q$ is the $q\times q$ identity matrix so that $\mathbf{Z}$ is a $Kq \times q$ matrix, and define
\begin{align*}
    \mathbf{X} = \big( \mathbf{I}_q \otimes (X', (Y_{t^*-1} - Y_{t^*-2}) \big)
\end{align*}
which is a $q \times q(K_X+1)$ matrix.  Finally, define the vector of parameters
\begin{align*}
    \gamma = (\beta_{t^*}^{*'}, F_{t^*}^*, \ldots, \beta_{\mathcal{T}}^{*'}, F_{\mathcal{T}}^*)'
\end{align*}
which is a $q(K_X+1) \times 1$ vector. 

Then, the vector of moment conditions coming from \Cref{eqn:momcond1} is equivalent to
\begin{align} \label{eqn:EZY}
  \E[\mathbf{Z}\Delta Y^{post,t^*-2}] = \E[\mathbf{ZX}]\gamma
\end{align}
We make the following assumption.
\begin{assumption}[Relevance] \label{ass:relevance}
  The matrix $\E[\mathbf{ZX}]$ has full rank.
\end{assumption}

\Cref{ass:relevance} corresponds to the relevance condition discussed above that the covariates whose effects do not change over time are correlated with $(Y_{t^*-1} - Y_{t^*-2})$ (or equivalently with $\lambda$) after controlling for other covariates $X$.

Next, we state our main identification result.
\begin{theorem} \label{thm:1}
  Under \Cref{ass:model,ass:sampling,ass:timing,ass:factors,ass:selection,ass:special-cov,ass:relevance}, $\beta^*_t$ and $F^*_t$ are identified for all $t=t^*,\ldots,\mathcal{T}$.  In particular, for any $Kq \times Kq$ positive definite weighting matrix $\mathbf{W}$,
  \begin{align*}
    \gamma = (\E[\mathbf{ZX}]' \mathbf{W} \E[\mathbf{ZX}])^{-1} \E[\mathbf{ZX}]' \mathbf{W} \E[\mathbf{Z}\Delta Y^{post,t^*-2}]
  \end{align*}
  In addition, $ATT_t$ is identified for all $t=t^*,\ldots,\mathcal{T}$, and it is given by
  \begin{align*}
    ATT_t = \E[Y_t - Y_{t^*-2} | D=1] - \Big(\E[X'|D=1]\beta^*_t + F^*_t \E[Y_{t^*-1} - Y_{t^*-2} | D=1] \Big)
  \end{align*}
\end{theorem}

\begin{proof}
For identifying $(\beta_t^*, F_t^*)$ for $t=t^*,\ldots,\mathcal{T}$, the result holds immediately from \Cref{eqn:EZY} under the conditions in the theorem.  The result for $ATT_t$ holds because
\begin{align*}
  ATT_t &= \E[Y_t(1) | D=1] - \E[Y_t(0)|D=1] \\
  &= \E[Y_t(1) - Y_{t^*-2}(0) | D=1] - \E[Y_t(0) - Y_{t^*-2}(0) | D=1] \\
  &= \E[Y_t - Y_{t^*-2} | D=1] - \E\left[X'\beta_t^* + F_t^* (Y_{t^*-1} - Y_{t^*-2}) | D=1\right]
\end{align*}
where the second equality holds by adding and subtracting $\E[Y_{t^*-2}(0)|D=1]$, and the third equality holds by \Cref{eqn:2,eqn:3} (and using that the coefficients on $W$ do not vary over time) and by rewriting potential outcomes in terms of their observed counterparts.  Then, the result holds because $\beta_t^*$ and $F_t^*$ are identified.
\end{proof}

In the expression for $ATT_t$ in \Cref{thm:1}, the first term is the mean path of outcomes that individuals in the treated group experienced between period $t^*-2$ and $t$.  The second term is the mean path of outcomes that individuals in the treated group would have experienced if they had not participated in the treatment and comes from the interactive fixed effects model for untreated potential outcomes in \Cref{ass:model}.

\begin{remark}
  We discuss two alternative approaches to identification in \Cref{app:alternative-approaches} in the Supplementary Appendix.  The first idea is to assume that the time varying unobservables, $U_{t}$, in the model in \Cref{ass:model} are uncorrelated over time; this sort of idea is common especially in the literature on factor models.  A recent paper that uses this sort of idea is \citet{imbens-kallus-mao-2021}.  Second, we consider the case with time varying covariates and using covariates in other periods as instruments.  Relative to our main approach, the main advantage of both of these approaches is that they do not require finding a covariate whose effect does not change over time.  On the other hand, they introduce additional complications and additional assumptions that often make these approaches less appealing in applications.  See \Cref{app:alternative-approaches} in the Supplementary Appendix for an expanded discussion of these issues.
\end{remark}

\begin{remark} \label{rem:pre-att}
    The above arguments apply for identifying $ATT_t$ for $t=t^*,\ldots,\mathcal{T}$ (i.e., post-treatment time periods).  In many applications, it may be useful to identify/estimate pseudo-ATTs in pre-treatment periods with the idea of pre-testing (i.e., testing that treatment effects are equal to 0 in pre-treatment periods).  In particular, define $b(t) := \begin{cases}
        t^*-2 & \textrm{if } t \geq t^* \\
        t-2 & \textrm{if } t < t^*
    \end{cases}$
    which allows the ``base period'' (the period that outcomes are differenced with respect to) to vary based on whether or not $t$ is post-treatment or pre-treatment.  In post-treatment periods $b(t)$ is equal to $t^*-2$ so that the base period is the same as has been used so far throughout this section.  In pre-treatment periods, $b(t)$ is equal to $t-2$ so that the base period is the period two periods before the current period.  Next, define $\Delta Y^{aug} := (Y_{3} - Y_{b(3)}, \ldots, Y_{\mathcal{T}} - Y_{b(\mathcal{T})})'$.  Similarly, define $\mathbf{Z}^{aug} := ( \mathbf{I}_{\mathcal{T}-2} \otimes (1-D)Z)$,  $\mathbf{X}^{aug}$ be the $\mathcal{T}-2 \times (\mathcal{T}-2)(K_X+1)$ block diagonal matrix with elements along the diagonal given by  $\big(X',(Y_{b(t)+1} - Y_{b(t)})\big)$, and $\gamma^{aug} := (\beta_3^{*'}, F_3^*, \ldots, \beta_{\mathcal{T}}^{*'}, F_{\mathcal{T}}^*)'$.  Then under analogous exogeneity and relevance conditions as above and using essentially the same arguments, one can show that, for some positive definite weighting matrix $\mathbf{W}$, $\gamma^{aug}=(\E[\mathbf{Z}^{aug}\mathbf{X}^{aug}]'\mathbf{W}\E[\mathbf{Z}^{aug} \mathbf{X}^{aug}])^{-1} \E[\mathbf{Z}^{aug}\mathbf{X}^{aug}]'\mathbf{W}\E[\mathbf{Z}^{aug} \Delta Y^{aug}]$.  Given this expression, $ATT_t = \Big(\E[Y_t | D=1] - \E[Y_{b(t)}|D=1]\Big) - \Big(\E[X'|D=1]\beta^*_t + F^*_t (\E[Y_{b(t)+1} - Y_{b(t)} | D=1]) \Big)$ which is identified for $t=3,\ldots,\mathcal{T}$.  For $t < t^*$, one can pre-test the interactive fixed effects model for untreated potential outcomes by testing that $ATT_t=0$.
\end{remark}

\begin{example} \label{ex:2} Under the same conditions as above, consider the case with exactly three time periods, where $W$ is a binary variable, and there are no other covariates in the model for untreated potential outcomes.

Straightforward calculations show that
\begin{align*}
  F_3^* = \frac{\E[Y_3 - Y_1|W=1,D=0] - \E[Y_3 - Y_1|W=0,D=0]}{\E[Y_2-Y_1|W=1,D=0] - \E[Y_2 - Y_1|W=0,D=0]}
\end{align*}
and
\begin{align*}
  \theta_3^* = \E[Y_3 - Y_1|W=0,D=0] - F_3^* \E[Y_2 - Y_1|W=0,D=0]
\end{align*}
which implies that
\begin{align*}
  ATT_3 = \E[Y_3 - Y_1|D=1] - \Big(\theta_3^* + F_3^* \E[Y_2 - Y_1|D=1]\Big)
\end{align*}
For $F_3^*$, notice that the difference in the numerator effectively differences out the time fixed effects (these are common across different values of $W$).  And, in particular, $\E[Y_3-Y_1|W=1,D=0] - \E[Y_3-Y_1|W=0,D=0] = (\E[\lambda|W=1,D=0] - \E[\lambda|W=0,D=0])(F_3 - F_1)$.  Thus, any differences in the numerator are due to $F_t$ and differences in the distribution of $\lambda$ among untreated individuals with $W=1$ relative to untreated individuals with $W=0$.  The same sort of argument implies that the denominator is equal to $(\E[\lambda|W=1,D=0] - \E[\lambda|W=0,D=0])(F_2 - F_1)$.  Thus, dividing the numerator by the denominator cancels the term involving $\lambda$ and is equal to $F_3^*$.    However, notice that this strategy would not be available without exploiting the time invariance of the effect of the covariate $W$.  Once $F_3^*$ has been identified, it is straightforward to recover $\theta_3^*$ and $ATT_3$.
\end{example}

\section{Consequences of Ignoring Interactive Fixed Effects}

\paragraph{Difference in Differences} \

The most common approach to identifying the effect of participating in a binary treatment when a researcher has access to repeated observations over time is difference in differences.  Difference in differences is very closely related to two-way fixed effects models for untreated potential outcomes; i.e.,
\begin{align} \label{eqn:twfe}
  Y_{it}(0) = \theta_t + \xi_i + U_{it}
\end{align}
Notice that the model in \Cref{eqn:twfe} is a special case of the model in \Cref{ass:model} --- when the effect of $\lambda_i$ does not vary over time, it can be absorbed into the individual fixed effect.  Under the condition that $\E[U_t | D=1] = \E[U_t | D=0]$ for all $t=1,\ldots,\mathcal{T}$ (which is a standard condition and one that holds under our conditions for interactive fixed effects models), it follows that
\begin{align} \label{eqn:twfe-id}
  \E[ \Delta Y_t(0) | D=1] = \E[\Delta Y_t(0) | D=0]
\end{align}
for all time periods $t=2,\ldots,\mathcal{T}$.  The condition in \Cref{eqn:twfe-id} is called the parallel trends assumption.  It says that the path of outcomes that individuals in the treated group would have experienced if they had not participated in the treatment is the same as the path of outcomes that individuals in the untreated group actually experienced.  It also immediately follows that\footnote{In this section, we focus on the $ATT$ in period $t^*$, but similar arguments would also apply in other time periods.}
\begin{align} \label{eqn:twfe-att}
  ATT_{t^*} = \E[\Delta Y_{t^*} | D=1] - \E[\Delta Y_{t^*} | D=0]
\end{align}
However, if the model in \Cref{ass:model} is correct instead of the model in \Cref{eqn:twfe}, the condition in \Cref{eqn:twfe-id} no longer holds.  In particular,
\begin{align*}
  \E[\Delta Y_{t^*}(0) | D=d] = (\theta_{t^*}-\theta_{t^*-1}) + (F_{t^*} - F_{t^*-1})\E[\lambda| D=d]
\end{align*}
and, therefore, the condition in \Cref{eqn:twfe-id} does not, in general, hold.  This follows because, although $\theta_t$ is common across the treated and untreated groups, $\E[\lambda|D=d]$ varies across groups.  And the second term will differ across groups unless either $\E[\lambda|D=1]=\E[\lambda|D=0]$ (in this case, the interactive fixed effects term can effectively be absorbed into the time fixed effect) or if $F_{t^*}=F_{t^*-1}$ (in this case, the interactive fixed effect can be absorbed into the individual fixed effect).  But these are both cases where the interactive fixed effect model reduces to the model in \Cref{eqn:twfe}.

Moreover, we can develop an expression for the bias that results from using a difference in differences approach when the actual model for untreated potential outcomes is the interactive fixed effects model.  Let $ATT^{DID}_{t^*}$ denote the expression on the right hand side of \Cref{eqn:twfe-att} but allowing for the possibility that the parallel trends assumption does not hold (so that, in this case, $ATT^{DID}_{t^*}$ may not be equal to $ATT_{t^*}$).  In this case,
\begin{align*}
  ATT^{DID}_{t^*} - ATT_{t^*} &= \E[\Delta Y_{t^*} | D=1] - \E[\Delta Y_{t^*} | D=0] - ATT_{t^*} \\
                 &= \underbrace{\big(\E[\Delta Y_{t^*}(1) | D=1] - \E[\Delta Y_{t^*}(0)|D=1]\big)}_{=ATT_{t^*}} \\
                 & \hspace{50pt} + \big(\E[\Delta Y_{t^*}(0) | D=1] - \E[\Delta Y_{t^*}(0) | D=0]\big) - ATT_{t^*} \\
                 &= (F_{t^*} - F_{t^*-1}) \big(\E[\lambda|D=1] - \E[\lambda|D=0]\big)
\end{align*}
where the first equality holds from the definition of $ATT^{DID}_{t^*}$, the second equality plugs in potential outcomes for their observed counterparts and adds and subtracts $\E[\Delta Y_{t^*}(0)|D=1]$, and the third equality holds by plugging in the interactive fixed effects model for untreated potential outcomes.  Thus, in general, wrongly imposing the parallel trends assumption instead of the interactive fixed effects model for untreated potential outcomes will lead to estimators that do not converge to the true $ATT_{t^*}$.  This bias is likely to be particularly severe in cases where $F_{t^*}$ is much different from $F_{t^*-1}$ (i.e., the effect of $\lambda$ is changing substantially over time) as well as when $\E[\lambda|D=1]$ is much different from $\E[\lambda|D=0]$ (i.e., the mean of $\lambda$ is much different between the treated and untreated groups).

\paragraph{Models with individual-specific linear trends} \ 

In applied work, it is very common that, when researchers suspect that the parallel trends assumption mentioned above is violated, to estimate a linear trends model.  These sorts of models (as well as more general versions) are considered in, for example, \citet{heckman-hotz-1989,wooldridge-2005,mora-reggio-2019}.  This sort of approach is motivated by the following model for untreated potential outcomes (as in the previous section, continue to consider the case without covariates)
\begin{align} \label{eqn:lineartrends}
  Y_{it}(0) = \theta_t + \xi_i + \lambda_i t + U_{it}
\end{align}
The model in \Cref{eqn:lineartrends} is a special case of the model in \Cref{ass:model} when $F_t$ is equal to $t$.  Let $C_t$ generically represent some variable that depends on time, and define the operator $\Delta^2 C_t = \Delta C_t - \Delta C_{t-1} = C_t - 2 C_{t-1} + C_{t-2}$.  Then, the model in \Cref{eqn:lineartrends} implies the following 
\begin{align} \label{eqn:lineartrends-id}
  \E[\Delta^2 Y_t(0) | D=1] = \E[\Delta^2 Y_t(0) | D=0]
\end{align}
for all time periods $t=3, \ldots, \mathcal{T}$. This further implies, after some straightforward calculations, that
\begin{align} \label{eqn:lineartrends-att}
  ATT_{t^*} = \E[\Delta^2 Y_{t^*} | D=1] - \E[\Delta^2 Y_{t^*} | D=0]
\end{align}
There are a couple of potential issues with the linear trends model.  First, conceptually, the reason to use this sort of model is the same as the interactive fixed effects model in \Cref{ass:model} --- particularly, there is some time invariant variable (i) that is unobserved and (ii) whose effect varies over time.  There are some notable drawbacks to this sort of model though.  In particular, if the researcher observed $\lambda_i$, under conditions (i) and (ii) mentioned above, it would be most natural to condition on it in the model and allow for its effect to vary over time.  This is exactly how the interactive fixed effects model treats this situation.  On the other hand, the linear trends model allows for the effect of $\lambda_i$ to change over time, but only in a very restrictive way.  In particular, if $\lambda_i$ were observed, it would be very unusual to impose a linear trend sort of specification and attach it to an observed covariate.

It is also helpful to consider the resulting bias in $ATT_{t^*}$ from incorrectly imposing the linear trends model.  When the model in \Cref{ass:model} is correct rather than \Cref{eqn:lineartrends}, the condition in \Cref{eqn:lineartrends-id} will not hold in general.  In particular, under the interactive fixed effects model
\begin{align*}
  \E[\Delta^2 Y_{t^*}(0) | D=d] = \Delta^2\theta_{t^*} + \Delta^2 F_{t^*} \E[\lambda|D=d]
\end{align*}
Thus, it is immediately clear that the condition in \Cref{eqn:lineartrends-id} does not hold except in special cases; $\Delta^2 \theta_{t^*}$ is common across groups but the second term, in general, varies across the treated and untreated group.  As mentioned above, linear trends is a special case of the interactive fixed effects model that we consider --- the interactive fixed effects model and the linear trends model coincide when $F_{t^*} - F_{t^*-1} = F_{t^*-1} - F_{t^*-2}$.  In this case, the second term is equal to 0 and therefore the condition in \Cref{eqn:lineartrends-id} holds.  The other case where \Cref{eqn:lineartrends-id} holds is when $\E[\lambda|D=1] = \E[\lambda|D=0]$, but, if the mean of $\lambda$ is the same across groups, there is no reason to use a linear trends model or interactive fixed effects model --- the term involving $\lambda$ will be absorbed into the time fixed effect.

Using similar arguments, we can also quantify the bias from incorrectly imposing a linear trends model when the interactive fixed effects model is correct.  Let $ATT^{LT}_{t^*}$ denote the expression on the right hand side of \Cref{eqn:lineartrends-att} but allowing for the possibility that the linear trends model in \Cref{eqn:lineartrends} is not correctly specified (as for the DID case above, this allows for the possibility that $ATT^{LT}_{t^*}$ is not equal to the $ATT_{t^*}$).  Then,
\begin{align*}
  ATT^{LT}_{t^*} - ATT_{t^*} &= \big(\E[\Delta^2 Y_{t^*} | D=1] - \E[\Delta^2 Y_{t^*} | D=0]) - ATT_{t^*} \\
                 &= \underbrace{\big( \E[\Delta^2 Y_{t^*}(1) | D=1] - \E[\Delta^2 Y_{t^*}(0) | D=1]\big)}_{=ATT_{t^*}} \\
                 & \hspace{50pt} + \big(\E[\Delta^2 Y_{t^*}(0) | D=1] - \E[\Delta^2 Y_{t^*}(0) | D=0]\big) - ATT_{t^*}\\
                 & = \E[\Delta^2 Y_{t^*}(0) | D=1] - \E[\Delta^2 Y_{t^*}(0) | D=0] \\
                 &= \Delta^2 F_{t^*} \big( \E[\lambda|D=1] - \E[\lambda|D=0] \big)
\end{align*}
where the first equality plugs in the definition of $ATT^{LT}_{t^*}$, the second equality plugs in potential outcomes for observed outcomes and adds and subtracts $\E[\Delta^2 Y_{t^*}(0) | D=1]$, the third equality cancels the terms involving $ATT_{t^*}$, and the fourth equality plugs in the interactive fixed effects model for untreated potential outcomes.  This expression demonstrates that there can be large biases from using a linear trends model instead of interactive fixed effects model in two cases:  (i) when $F_{t^*} - F_{t^*-1}$ is much different from $F_{t^*-1} - F_{t^*-2}$, or (ii) when $\E[\lambda|D=1]$ is much different from $\E[\lambda|D=0]$.

\section{Estimation and Inference}

In this section, we propose a two step procedure for estimating $ATT_t$.  In the first step, we estimate all the parameters $\gamma$ as well as $\E[Y_t - Y_{t^*-2} | D=1]$, $\E[X|D=1]$, $\E[Y_{t^*-1}-Y_{t^*-2}|D=1]$, and $p:=\P(D=1)$.  Then, we plug these into the expression for $ATT_t$ in \Cref{thm:1}.  In the second part of this section, we derive the asymptotic distribution of our estimator of $ATT_t$ and propose an approach to conduct inference based on the multiplier bootstrap.

As a first step, it is helpful to re-write the expression for $ATT_t$ in \Cref{thm:1}  as
\begin{align} \label{eqn:attv2}
  ATT_t &= \frac{\E[D(Y_t - Y_{t^*-2})]}{p} - \frac{\E[DX']}{p}\beta_t^* - F_t^* \frac{\E[D(Y_{t^*-1} - Y_{t^*-2})]}{p}
\end{align}
which just converts the conditional moments in \Cref{thm:1} into equivalent unconditional moments.  This expression suggests the following estimator for $ATT_t$
\begin{align} \label{eqn:att-hat}
    \widehat{ATT}_t = \hat{p}^{-1} \left( \frac{1}{n} \sum_{i=1}^n D_i (Y_{it} - Y_{it^*-2}) -  \frac{1}{n}\sum_{i=1}^n D_i X_i' \hat{\beta}_t^* - \hat{F}_t^* \frac{1}{n}\sum_{i=1}^n D_i (Y_{it^*-1} - Y_{it^*-2})\right)
\end{align}

Next, we provide the limiting distribution for our estimators of $ATT_t$.  Before doing that, we define some additional notation.  First, define $V_i := \Delta Y_i^{post,t^*-2} - \mathbf{X}_i \gamma$.  Next, define $\bm{\Xi}_t := [\mathbf{0}_{K_X+1,(t-t^*)(K_X+1)}, \mathbf{I}_{K_X+1}, \mathbf{0}_{K_X+1, (\mathcal{T}-t)(K_X+1)}]$ where $\mathbf{0}_{\ell_1,\ell_2}$ denotes an $\ell_1 \times \ell_2$ matrix of zeros, and so that $(\beta_t^{*'}, F_t^*)' = \bm{\Xi}_t \gamma$ for any $t \in \{t^*,\ldots,\mathcal{T}\}$, and define $Y_i := (Y_{it^*-2}, Y_{it^*-1}, \ldots, Y_{i\mathcal{T}})$.  Next, define
\begin{align*}
    \varphi^P(D_i) &:= D_i - \E[D] \\ 
    \varphi^M_t(D_i, X_i, Y_i) &:= \begin{pmatrix}
      D_i(Y_{it} - Y_{it^*-2}) - \E[D(Y_t - Y_{t^*-2})] \\
      D_i X_i - \E[DX] \\
      D_i (Y_{it^*-1} - Y_{it^*-2}) - \E[D (Y_{t^*-1}- Y_{t^*-2})]
    \end{pmatrix} \\
    \varphi^{GMM}_t(D_i, Z_i, Y_i) &:= \mathbf{\Xi}_t (\E[\mathbf{ZX}]' \mathbf{W} \E[\mathbf{ZX}])^{-1} \E[\mathbf{ZX}]' \mathbf{W} \mathbf{Z}_i V_i
\end{align*}
where $\varphi^P$ is a scalar, $\varphi^M_t$ is a $(K_X+2)$-dimensional vector, and $\varphi^{GMM}_t$ is a $(K_X+1)$-dimensional vector.  Further, define

\begin{align*}
    \psi_t^A(D_i) &:= - \frac{ATT_t}{p^2} \varphi^P(D_i) \\
    \psi^B_t(D_i,X_i,Y_i) &:= p^{-1} \varphi^M_t(D_i, X_i, Y_i)' \begin{pmatrix} 1 \\ \beta_t^* \\ F_t^* \end{pmatrix} \\
    \psi_t^C(D_i, Z_i, Y_i) &:= p^{-1} \begin{pmatrix} \E[DX] \\ \E[D(Y_{t^*-1} - Y_{t^*-2})] \end{pmatrix}' \varphi^{GMM}_t(D_i, Z_i, Y_i)
\end{align*}
Finally, define $\psi_t(D_i, Z_i, Y_i) := \psi_t^A(D_i) + \psi_t^B(D_i, X_i, Y_i) + \psi_t^C(D_i, Z_i, Y_i)$, $\Psi(D_i,Z_i,Y_i) := \big(\psi_{t^*}(D_i,Z_i,Y_i), \ldots, \psi_{\mathcal{T}}(D_i,Z_i,Y_i)\big)'$, $\widehat{ATT} := (\widehat{ATT}_{t^*}, \ldots, \widehat{ATT}_{\mathcal{T}})'$, and $ATT := (ATT_{t^*}, \ldots, ATT_{\mathcal{T}})'$.

We make the following assumption
\begin{assumption} \label{ass:asymptotic}
  $\E\|Y\|^4 < \infty$, $\E\|Z\|^4 < \infty$, $\mathbf{\Omega} := \E[\mathbf{Z}V V' \mathbf{Z}']$ is positive definite, and $\hat{\mathbf{W}} \xrightarrow{p} \mathbf{W}$ which is positive definite.
\end{assumption}
\Cref{ass:asymptotic} is a standard condition for establishing the limiting distribution of GMM estimators.

The next result provides an asymptotically linear representation for our estimator of $ATT_t$ and the joint limiting distribution of $ATT_t$ for $t=t^*,\ldots,\mathcal{T}$.

\begin{theorem} \label{thm:asy-lin} Under  \Cref{ass:model,ass:sampling,ass:timing,ass:factors,ass:selection,ass:special-cov,ass:relevance,ass:asymptotic} and for any $t \in \{t^*, \ldots, \mathcal{T}\}$,
\begin{align*}
    \sqrt{n}\big(\widehat{ATT}_t - ATT_t\big) = \frac{1}{\sqrt{n}} \sum_{i=1}^n \psi_t(D_i, Z_i, Y_i) + o_p(1)
\end{align*}
In addition,
\begin{align*}
    \sqrt{n}\big(\widehat{ATT} - ATT\big) \xrightarrow{d} N(0,\mathbf{\Sigma})
\end{align*}
where $\mathbf{\Sigma} = \E[\Psi(D,Z,Y) \Psi(D,Z,Y)']$.
\end{theorem}

The proof of \Cref{thm:asy-lin} is provided in \Cref{app:additional-proofs} in the Supplementary Appendix.  Given the result in \Cref{thm:asy-lin}, one can immediately proceed to conduct inference by estimating $\bm{\Sigma}$.   Instead of doing this, we conduct inference using a multiplier bootstrap procedure (see, for example, \citet{kline-santos-2012}) that involves perturbing the influence function.  In particular, for some iid $\upsilon_i$ that has mean zero, variance one, and finite third moment and that is drawn independently of the data (common examples include drawing $\upsilon_i$ from a standard normal distribution, or setting $\upsilon_i$ to be equal to $1$ or $-1$ each with probability $1/2$), we construct a bootstrap estimate by
\begin{align*}
    \widehat{ATT}^* = \widehat{ATT} + \frac{1}{n}\sum_{i=1}^n \upsilon_i \hat{\Psi}(D_i,Z_i,Y_i)
\end{align*}
where $\hat{\Psi}(D_i,Z_i,Y_i)$ is the estimated version of $\Psi(D_i,Z_i,Y_i)$ that replaces population parameters with their estimates.  Standard arguments show that, conditional on the original data, $\sqrt{n}(\widehat{ATT}^* - \widehat{ATT})$ follows the same limiting distribution as in \Cref{thm:asy-lin}.  Thus, for $t \in \{t^*,\ldots,\mathcal{T}\}$ one can construct standard errors for $\widehat{ATT}_t$ by $\hat{\bm{\Sigma}}_{tt}^{1/2}/\sqrt{n}$ where $\hat{\bm{\Sigma}}_{tt}^{1/2} := \big(q_{0.75}(t) - q_{0.25}(t)\big)/\big(z_{0.75} - z_{0.25}\big)$ and where $q_{\tau}(t)$ is the $\tau$-th sample quantile of $\sqrt{n}(\widehat{ATT}^*_t - \widehat{ATT}_t)$ (across bootstrap iterations) and $z_\tau$ is the $\tau$-th quantile of the standard normal distribution.  Relative to the nonparametric bootstrap, the multiplier bootstrap offers a number of practical advantages: (i) because it does not involve re-estimating the model at each bootstrap iteration, it is generally much faster; (ii) because it involves weights, it avoids cases where, in some bootstrap iterations, the nonparametric bootstrap may not be able to compute a bootstrap estimate.

\begin{remark} \label{rem:testing}
  In cases where a researcher would like to ``pre-test'' the identifying assumptions (see discussion in \Cref{rem:pre-att} for more details), similar arguments to the ones in \Cref{thm:asy-lin} can be used to establish the joint limiting distribution of $ATT_t$ in pre-treatment periods (or across all pre- and post-treatment periods) which immediately leads to tests of $ATT_t=0$ for all $t=3,\ldots,t^*-1$.
\end{remark}

\begin{remark}
  Although our main results consider the case with iid observations, it would also be straightforward to extend the result to account for clustering across individuals, allowing, for example, spatial correlations as long as the number of clusters is ``large.''  In that case, the asymptotic arguments would proceed with the number of clusters, rather than the number of individuals, going to infinity.  It is likely that various bootstrap procedures could be adapted to this case as well (see, for example, \citet{cameron-miller-2015}).
\end{remark}

\section{Extensions}

This section expands our identification arguments from the previous section to allow for (i) multiple interactive fixed effects (and, relatedly, provides an approach for selecting the number of interactive fixed effects), (ii) repeated cross sections data, (iii) testing that particular covariates have time invariant effects when additional pre-treatment time periods are available, and (iv) variation in treatment timing across individuals.

\subsection{Multiple Interactive Fixed Effects} \label{sec:multiple-ife}

The previous results dealt with the case with a single interactive fixed effect.  This section generalizes those arguments to cases where there may be more interactive fixed effects.

Consider the case where there are $R$ interactive fixed effects so that
\begin{align} \label{eqn:multiple-ifes}
  Y_{it}(0) &= \xi_i + \lambda_i'F_t + X_i'\beta_t + W_i'\alpha + U_{it}
\end{align}
where $\lambda_i = (\lambda_{i1}, \lambda_{i2}, \ldots, \lambda_{iR})'$ and $F_t = (F_{1t}, F_{2t}, \ldots, F_{Rt})'$ are each $R \times 1$ vectors, $X_i$ is a $K_X \times 1$ vector, and $W_i$ is a $K_W \times 1$ vector.  

Next, for any time period $t \geq (t^*-R)$, notice that
\begin{align} \label{eqn:mife-ld}
  Y_{it}(0) - Y_{it^*-R-1}(0) &= \lambda_i'(F_t - F_{t^*-R-1}) + X_i'(\beta_t - \beta_{t^*-R-1}) + (U_{it} - U_{it^*-R-1})
\end{align}
Next, define $\Delta Y^{(t_1, t_2)}(0) := Y_{t_2}(0) - Y_{t_1}(0)$, $\Delta \beta^{(t_1,t_2)} = \beta_{t_2} - \beta_{t_1}$, and $\Delta F^{(t_1,t_2)} = F_{t_2} - F_{t_1}$.  Further, define $\Delta Y^{pre,R}(0) := (\Delta Y^{(t^*-R-1,t^*-1)}, \ldots, \Delta Y^{(t^*-R-1,t^*-R)})'$ which is an $R$ dimensional vector of differences between all pre-treatment outcomes relative to the ``base period'' $t^*-R-1$.  Similarly, define $\bm{\Delta \beta}_{pre,R} := (\Delta \beta^{(t^*-R-1,t^*-1)}, \ldots, \Delta \beta^{(t^*-R-1,t^*-R)})'$ (which is an $R \times K_X$ matrix), $\bm{\Delta F}_{pre,R} := (\Delta F^{(t^*-R-1,t^*-1)}, \ldots, \Delta F^{(t^*-R-1,t^*-R)})'$ (which is an $R \times R$ matrix), and $\Delta U^{pre,R} := (\Delta U^{(t^*-R-1,t^*-1)}, \ldots, \Delta U^{(t^*-R-1,t^*-R)})'$.  We make the following assumption
\begin{assumption}[Factor Rank] \label{ass:factor-rank} $\bm{\Delta F}_{pre,R}$ has full rank.
\end{assumption}
This is the analogous condition to \Cref{ass:factors} in the case with a single interactive fixed effect.

Next, notice that we can write
\begin{align*}
  \Delta Y_i^{pre,R}(0) = \bm{\Delta F}_{pre,R} \lambda_i + \bm{\Delta \beta}_{pre,R} X_i + \Delta U_i^{pre,R}
\end{align*}
which, under \Cref{ass:factor-rank}, implies that
\begin{align} \label{eqn:multiple_ifes_lambda}
  \lambda_i = \bm{\Delta F}_{pre,R}^{-1} \Delta Y_i^{pre,R}(0) - \bm{\Delta F}_{pre,R}^{-1} \bm{\Delta \beta}_{pre,R} X_i - \bm{\Delta F}_{pre,R}^{-1} \Delta U_i^{pre,R}
\end{align}
Plugging this expression for $\lambda_i$ into \Cref{eqn:mife-ld}, notice that, for any $t \in \{t^*,\ldots,\mathcal{T}\}$,
\begin{align} \label{eqn:multiple_ifes_estimating_equation}
  Y_{it}(0) - Y_{it^*-R-1}(0) 
  &= X_i'\beta^*_t + \Delta Y_i^{pre,R}(0)'F^*_t + V_{it} 
\end{align}
where $\beta_t^* := \left( (\beta_t - \beta_{t^*-R-1}) - \big( \bm{\Delta F}_{pre,R}^{-1} \bm{\Delta \beta}_{pre,R} \big)'(F_t - F_{t^*-R-1}) \right)$, $F_t^* := \big(\bm{\Delta F}_{pre,R}^{-1} \big)' (F_t - F_{t^*-R-1})$, and $V_{it} := U_{it} - U_{it^*-R-1} - (F_t - F_{t^*-R-1})'\bm{\Delta F}^{-1}_{pre,R} \Delta U^{pre,R}_i$.

This expression in \Cref{eqn:multiple_ifes_estimating_equation} is the natural generalization of \Cref{eqn:2} to the case with multiple interactive fixed effects.  As in the earlier case, to recover $ATT_t$, we do not need to recover entire vectors of parameters such as $F_t$ or $\beta_t$, but rather we need only recover $\beta^*_t$ and $F^*_t$.  As before, the main complication is that, by construction, all the $R$ terms in $\Delta Y^{pre,R}(0)$ are endogenous.  Thus, a necessary condition for identifying $ATT_t$ in this case is that $K_W$ (the dimension of the time invariant covariates $W$ whose effects do not change over time) is greater than or equal to $R$.  In other words, to accommodate more interactive fixed effects, the researcher needs to be able to find more time invariant covariates whose effects do not change over time.  Additionally, the arguments above for differencing out the interactive fixed effects rely on $\Delta Y^{pre,R}(0)$ which involves untreated potential outcomes in all time periods from $t^*-R-1$ to $t^*-1$ --- thus, including $R$ interactive fixed effects requires having access to at least $R+1$ pre-treatment time periods (as well as at least $R+2$ total time periods).  Define $\mathbf{X}_M := \mathbf{I}_q \otimes (X',\Delta {Y^{pre,R}}')$, which is a $q\times q(K_X + R)$ matrix, and define $\Delta Y^{post,R} = (Y_{t^*}-Y_{t^*-R-1}, \ldots, Y_{\mathcal{T}} - Y_{t^*-R-1})'$ which is a $q \times 1$ vector.  We state the following result that generalizes \Cref{thm:1} to the case with multiple interactive fixed effects.

\begin{proposition} \label{prop:multiple-ifes} In the model in \Cref{eqn:multiple-ifes}, and under \Cref{ass:sampling,ass:selection,ass:special-cov,ass:factor-rank}, and under the additional conditions that (i) $\mathcal{T} \geq R+2$ and $t^* \geq R+2$, (ii) $K_W \geq R$, and (iii) $\E[\mathbf{Z}\mathbf{X}_M]$ has full rank, then $\beta^*_t$ and $F^*_t$ are identified for all $t=t^*,\ldots,\mathcal{T}$.  In particular, for any $Kq \times Kq$ positive definite weighting matrix $\mathbf{W}$,
  \begin{align*}
    \gamma = (\E[\mathbf{Z} \mathbf{X}_M]' \mathbf{W} \E[\mathbf{Z} \mathbf{X}_M])^{-1} \E[\mathbf{Z} \mathbf{X}_M]' \mathbf{W} \E[\mathbf{Z} \Delta Y^{post,R}]
\end{align*}
  In addition, $ATT_t$ is identified for all $t=t^*,\ldots,\mathcal{T}$, and it is given by
  \begin{align*}
    ATT_t = \E[Y_t - Y_{t^*-R-1} | D=1] - \left(\E[X'|D=1]\beta^*_t +  \E[\Delta {Y^{pre,R}}' | D=1] F^*_t\right)
\end{align*}
\end{proposition}

The proof of \Cref{prop:multiple-ifes} is provided in \Cref{app:additional-proofs} in the Supplementary Appendix and follows from arguments that are similar to the ones in \Cref{thm:1}.

\subsubsection*{Selecting the Number of Interactive Fixed Effects}

So far, we have considered the case where the researcher knows the number of interactive fixed effects.  This may not always be the case though, and it is common in the interactive fixed effects literature to choose the number of interactive fixed effects in a data driven way.  Suppose that the researcher chooses among $R=0,\ldots,R^*$ interactive fixed effects and that $K_W \geq R^*$ so that there are enough covariates whose effects do not change over time to identify the parameters of the model for untreated potential outcomes.  Building on existing work, we suggest three approaches to choosing the number of interactive fixed effects.  First, we consider choosing the number of interactive fixed effects according to which one minimizes the Bayesian Information Criteria, $BIC = J - \log(n)(K-K_X-R+1)$ where $J = n \bar{g}_n' \hat{\mathbf{W}} \bar{g}_n$ where $\bar{g}_n := n^{-1}\sum_{i=1}^n \mathbf{Z}_i \hat{V}_i$ and where $\hat{V}_i$ is the $q\times 1$ vector of residuals from estimating the interactive fixed effects model for untreated potential outcomes in \Cref{eqn:multiple_ifes_estimating_equation} in post-treatment time periods.\footnote{The expression for BIC only uses post-treatment periods, but when there are ``extra'' pre-treatment time periods, it is straightforward to include those periods.}  Related approaches are proposed in \citet{cragg-donald-1997,bai-ng-2002,ahn-lee-schmidt-2013,robertson-sarafidis-2015,juodis-sarafidis-2020}.  

Second, we consider leave-one-out cross validation using untreated observations.  In particular, in a particular post-treatment period $t \geq t^*$, for a particular individual in the untreated group, we estimate $(\hat{\beta^*}_{-i,t}',\hat{F^*}_{-i,t})'$ where the notation indicates the estimates of $\beta^*_t$ and $F^*_t$ that use all observations except observation $i$.  Then, we compute $CV_U(t) = \sum_{i \in \mathcal{U}} \Big( (Y_{it} - Y_{it^*-R-1}) - (X_i'\hat{\beta^*}_{-i,t} - \Delta Y_i^{pre,R}(0)' \hat{F^*}_{-i,t} ) \Big)^2$ where $\mathcal{U}$ denotes the set of untreated observations.  And we choose $R \in \{0, \ldots, R^*\}$ that minimizes $CV_U = \sum_{t=t^*}^{\mathcal{T}} CV_U(t)$.  A similar approach is proposed in \citet{liu-wang-xu-2021}.   

Finally, we consider a cross-validation procedure based on choosing the number of interactive fixed effects using pre-treatment observed outcomes for the treated group.  This is conceptually similar to the approach considered in \citet{xu-2017}.  In particular, for some time period $2+R^* \leq t < t^*$ (these are the common pre-treatment periods where pseudo-treatment effect parameters can be estimated), compute $CV_T(t) = \sum_{i \in \mathcal{D}} \Big( (Y_{it} - Y_{it-R-1}) - (X_i'\hat{\beta}^*_t - \Delta Y^{pre,R}_i(0)' \hat{F}^*_t ) \Big)^2$ where $\mathcal{D}$ denotes the set of treated observations.  Then, we choose $R \in \{0,\ldots,R^*\}$ that minimizes $CV_T = \sum_{t=2+R^*}^{t^*-1} CV_T(t)$.

\subsection{Repeated Cross Sections} \label{sec:repeated-cross-sections}

It is straightforward to extend our main identification arguments to the case when the researcher has access to repeated cross sections data rather than panel data.  We appear to be the first paper to consider interactive fixed effects models with repeated cross sections data.  That our approach can be extended to this case is in parallel with the literature on difference in differences which can often be implemented with repeated cross sections data.  In our case, these arguments work because of (i) our differencing argument combined with the linearity of expectations and (ii) our focus on time invariant covariates.  In our view, having results available for the case with repeated cross sections data is useful because repeated cross sections are often available, typically have larger sample sizes, and have fewer issues with attrition.  These types of considerations are particularly important in policy evaluation applications which often exploit variation in local policies where typical panel datasets can frequently be too small to be able to be used once a researcher focuses on particular local policies.  In this section, we let $T_i$ denote the time period when individual $i$ is observed.  With repeated cross sections data, we replace \Cref{ass:sampling} with the following assumption,
\begin{uassumption}{RC} [Repeated Cross Sections Sampling] \label{ass:sampling-rc} Conditional on $T=t$, the data are iid from the distribution of $(Y_t, Z, D)$ for all $t=1,\ldots,\mathcal{T}$, and the joint distribution of $(Y_1,\ldots,Y_{\mathcal{T}},Z,D)$ is independent of $T$.
\end{uassumption}
\Cref{ass:sampling-rc} implies that we observe draws from a distribution which may not correspond to the population distribution of outcomes, covariates, and treatments.  That is, within each period, we observe iid draws from some underlying population, but we only observe outcomes for individuals in the particular period when they are observed in the sample and the sample sizes can vary across different periods.  In particular, we observe draws from the mixture distribution
\begin{align*}
  F_M(y,z,d,t) = \sum_{t=1}^{\mathcal{T}} \pi_t F_{Y_t,Z,D|T}(y,z,d|t)
\end{align*}
where $\pi_t := \P(T=t)$.  The second part of \Cref{ass:sampling-rc} rules out compositional changes in outcomes, covariates, and treatment status across time periods (see \citet{hong-2013} for a more extended discussion of compositional changes though we note here that this type of condition is common in the literature on DID with repeated cross-sections; e.g., \citet{abadie-2005,santanna-zhao-2020}).  Let expectations with respect to the mixture distribution be denoted by $\E_M[\cdot]$ (and population expectations continue to be denoted by $\E[\cdot]$).  

Notice that all the moment conditions in \Cref{eqn:momcond1} continue to hold, but, under \Cref{ass:sampling-rc}, we do not have direct sample analogues of these moment conditions.  First, for $t=1,\ldots,\mathcal{T}$, let $T_t = \indicator{T = t}$.  Then, notice that for expectations of variables in a particular time period such as $\E[Y_t]$, under \Cref{ass:sampling-rc} we can write $\E[Y_t] = \E_M[Y | T_t=1] = \E_M[T_t Y / \pi_t]$; and for expectations of variables that are time invariant such as $\E[X]$, under \Cref{ass:sampling-rc}, we can write $\E[X] = \E_M[X]$. Then, exploiting the linearity of expectations, we can re-write \Cref{eqn:momcond1} in terms of expectations from the mixture distribution as 
\begin{align} \label{eqn:rc-moment-conditions}
  0 = \E_M\left[ (1-D) Z \left( \frac{T_t}{\pi_t} Y - \frac{T_{t^*-2}}{\pi_{t^*-2}} Y -  X'\beta^*_t - F^*_t \left( \frac{T_{t^*-1}}{\pi_{t^*-1}}Y - \frac{T_{t^*-2}}{\pi_{t^*-2}} Y\right) \right) \right] \quad \textrm{for } t=t^*,\ldots,\mathcal{T}
\end{align}

Given the modified moment conditions above, identification of $ATT_t$ follows from essentially the same arguments as for the case with panel data.  We briefly state these arguments here for completeness.  Next, let 
\begin{align*}
  \Delta Y^{RC} = \left( \frac{T_{t^*}}{\pi_t} Y - \frac{T_{t^*-2}}{\pi_{t^*-2}}Y, \ldots,  \frac{T_{\mathcal{T}}}{\pi_{\mathcal{T}}} Y - \frac{T_{t^*-2}}{\pi_{t^*-2}} Y \right)'
\end{align*}
which is a $q \times 1$ vector.  The $Kq \times q$ instrument matrix $\mathbf{Z}$ is exactly the same as in the case with panel data.  Next, define
\begin{align*}
  \mathbf{X}_{RC} = \left(\mathbf{I}_{q} \otimes \left(X', \left(\frac{T_{t^*-1}}{\pi_{t^*-1}} Y - \frac{T_{t^*-2}}{\pi_{t^*-2}} Y \right)\right)\right)
\end{align*}
which is a $q \times q(K_X+1)$ matrix.  Finally, we state an identification result for $ATT_t$ when repeated cross sections data is available.

\begin{proposition} \label{prop:repeated-cross-sections}
  Under Assumptions \ref{ass:model}, \ref{ass:sampling-rc}, and \ref{ass:timing} to \ref{ass:relevance}, $\beta^*_t$ and $F^*_t$ are identified for all $t=t^*,\ldots,\mathcal{T}$.  In particular, for any $Kq \times Kq$ positive definite weighting matrix $\mathbf{W}$,
  \begin{align*}
    \gamma = (\E_M[\mathbf{ZX}_{RC}]' \mathbf{W} \E_M[\mathbf{ZX}_{RC}])^{-1} \E_M[\mathbf{ZX}_{RC}]' \mathbf{W} \E_M[\mathbf{Z} \Delta Y^{RC}]
  \end{align*}
  In addition, $ATT_t$ is identified for all $t=t^*,\ldots,\mathcal{T}$, and it is given by
  \begin{align*}
    ATT_t &= \E_M\left[\frac{T_t}{\pi_t} Y - \frac{T_{t^*-2}}{\pi_{t^*-2}} Y \Big| D=1\right]  \\ 
    & \hspace{50pt} - \Bigg\{\E_M[X'|D=1]\beta^*_t + F^*_t \left(\E_M\left[\frac{T_{t^*-1}}{\pi_{t^*-1}} Y  - \frac{T_{t^*-2} }{\pi_{t^*-2}} Y \Big| D=1\right]\right) \Bigg\}
  \end{align*}
\end{proposition}

The proof of \Cref{prop:repeated-cross-sections} is provided in \Cref{app:additional-proofs} in the Supplementary Appendix and follows using essentially the same arguments as in the proof of \Cref{thm:1}.

\subsection{Testing for Covariates Having Time Invariant Effects} \label{sec:time-invariant-effects-test}

Our identification arguments rely heavily on finding a covariate whose effect on untreated potential outcomes does not vary over time.  In many applications, it may be unclear whether a particular covariate has a time invariant effect on untreated potential outcomes or not.  For example, in our application on job displacement, it seems likely that the covariate AFQT score (which is a proxy for individual skill/ability) does not have a time varying effect on outcomes after accounting for unobserved heterogeneity.  But it is less clear if the other covariates, which are education and demographic characteristics, have time invariant effects.  Even for AFQT, it would be nice to have quantitative evidence that it does not have time varying effects on untreated potential outcomes.  

One idea here is a traditional over-identification test; i.e., when $K_W$ (the dimension of the covariates with time invariant effects) is greater than $R$ (the number of interactive fixed effects), one can jointly test the restrictions that the corresponding parameters do not vary over time.  This is straightforward to do and is a byproduct of the first-stage estimation.

Unfortunately, this strategy is not available in the leading case where $R=K_W=1$ (that is, there is one interactive fixed effect and one covariate whose effect does not change over time) because the model for untreated potential outcomes is exactly identified in this case.  A useful alternative approach is available in this case when an extra pre-treatment period is available in the data (i.e., when $t^* > 3$).  In that case, notice that the following additional moment conditions are available in periods $3, \ldots, t^*-1$,
\begin{align*}
    \E[ZV_t | D=1] = 0 \implies \E\left[ \begin{pmatrix} D \\ Z \end{pmatrix} V_t \right] = 0
\end{align*}
This is a $(K+1) \times 1$ vector of moment conditions and thus satisfies the necessary order condition in pre-treatment periods.  The next proposition shows that $\beta^*_t$ and $F^*_t$ are identified for $t=3,\ldots,t^*-1$ without requiring having access to a covariate whose effect does not vary over time.

\begin{proposition} \label{prop:covariate_pre_test} For $t \in \{ 3, \ldots, t^*-1\}$, under Assumptions \ref{ass:model}, \ref{ass:sampling}, \ref{ass:selection}, and under the additional conditions that $\mathcal{T} \geq 4$ and $t^* \geq 4$, that $F_{t-1} \neq F_{t-2}$, and that $\E\left[ \begin{pmatrix} D \\ Z \end{pmatrix} (Z', Y_{t-1} - Y_{t-2})\right]$ is positive definite, then
\begin{align*}
    \begin{pmatrix}
      \delta^*_t \\ F^*_t
    \end{pmatrix} = \E\left[ \begin{pmatrix} D \\ Z \end{pmatrix} (Z', Y_{t-1} - Y_{t-2})\right]^{-1} \E\left[ \begin{pmatrix} D \\ Z \end{pmatrix} (Y_t - Y_{t-2}) \right]
\end{align*}
where, along the lines of \Cref{rem:pre-att}, in pre-treatment periods we define $\delta_t^* = \delta_t - \delta_{t-2} - F_t^*(\delta_{t-1} - \delta_{t-2})$ and $F_t^* = (F_t - F_{t-2}) \Big/ (F_{t-1} - F_{t-2})$ (i.e., here, parameters indexed by $t-1$ and $t-2$ take the place of parameters indexed by $t^*-1$ and $t^*-2$ as was the case in post-treatment time periods).  
\end{proposition}
The proof of \Cref{prop:covariate_pre_test} is provided in \Cref{app:additional-proofs} in the Supplementary Appendix.  The key difference of \Cref{prop:covariate_pre_test} relative to \Cref{thm:1} is that this result does not require \Cref{ass:special-cov} --- that the effect of a particular covariate does not vary over time.  Further, note that the entire vector of parameters $\delta_t^*$ is identified in \Cref{prop:covariate_pre_test} in pre-treatment periods.  Moreover, given the definition of $\delta^*_t$, an implication of a particular covariate having time invariant effects is that $\delta^*_{jt} = 0$ for all $t=3,\ldots,t^*-1$.  This implies that one can therefore test for time invariant effects of a particular covariate in pre-treatment periods by estimating the model for untreated potential outcomes in pre-treatment periods using the result in \Cref{prop:covariate_pre_test} and then simply testing whether $\delta^*_{jt}=0$ for all $t=3,\ldots,t^*-1$.  Note that this is not a direct test that $\delta_{jt}$ does not vary over time in post-treatment time periods, but it is similar to pre-tests that are extremely common in the DID literature.  We also stress that it is likely not appropriate to \textit{choose} which covariates to use as those that have time invariant effects based on this kind of pre-test (see \citet{roth-2020} for related discussion on pre-testing in the context of difference in differences).  

\subsection{Staggered Treatment Adoption} \label{sec:staggered}

Next, we consider the case where treatment timing can potentially vary across different individuals.  We also address the issue of treatment anticipation in this section. These issues have been the subject of a number of recent papers in the difference in differences literature (e.g., \citet{chaisemartin-dhaultfoeuille-2020,goodman-2021,callaway-santanna-2021}).  Throughout this section, we make the following assumption.

\begin{assumption}[Staggered Treatment Adoption] \label{ass:staggered} For all $t=2, \ldots, \mathcal{T}$, $D_{it-1} = 1 \implies D_{it}=1$.
\end{assumption}

\Cref{ass:staggered} says that once an individual becomes treated, they remain treated.  This is a very common setup in applied work and common assumption in the econometrics literature.  It would apply when a particular location implements a policy and the policy remains in place; it also applies in applications where treatments are ``scarring'' in the sense that when an individual becomes treated, the researcher considers them to be treated in future periods as well even if they do not repeatedly participate in the treatment.  This is the case in our application on job displacement.

In this section, we slightly modify the notation used in the previous sections.  Because treatment timing can vary across individuals, we define an individual's ``group'' by the time period when they become treated; that is, for each individual, we define $G_i$ as the time period when an individual becomes treated, and (arbitrarily) we set $G_i=0$ for individuals that are not treated in any time period.  Under \Cref{ass:staggered}, knowing an individual's treatment timing implies that we know their entire path of participating in the treatment.  We continue to define an individual's untreated potential outcome by $Y_{it}(0)$ --- this is the outcome an individual would experience if they did not participate in the treatment in any time period.  But now we index treated potential outcomes by $g$; that is, $Y_{it}(g)$ is the outcome that an individual would experience in time period $t$ if they became treated in period $g$.  In all time periods, we observe $Y_{it} = Y_{it}(G_i)$ which is the potential outcome corresponding to the actual time period when individual $i$ became treated.  We make the following assumption.

\begin{assumption}[Anticipation] \label{ass:anticipation} There exists a $\tau \geq 0$, such that, $Y_{it}(G_i) = Y_{it}(0)$ for all $t \leq G_i -\tau-1$.
\end{assumption}

\Cref{ass:anticipation} allows for individuals to anticipate participating in the treatment and therefore have their outcomes affected before the treatment actually takes place.  However, it says that, if we move back in time by enough periods, then the treatment does not have effects in those periods.  Let $\mathcal{G} := \textrm{support}(G)\setminus \{0\} \subseteq \{3,\ldots,\mathcal{T}\}$ denote the set of groups that ever participate in the treatment (and where the notation indicates that we continue to consider the case where there are at least two pre-treatment time periods available for all individuals).  In the presence of anticipation, it is also useful to define the subset of groups that are treated late enough so that at least two periods of outcomes that are not affected by the treatment (either directly or through anticipation) are available.  In particular, let $\mathcal{G}_\tau = \{ g \in \mathcal{G} : g \geq 3+\tau\}$;\footnote{In some applications it may also be desirable to limit $\mathcal{G}_{\tau}$ to groups that are treated early enough (as well as early enough time periods) so that no anticipation effects could be occurring among the comparison group.  For simplicity (as well as following standard approaches in most applications), we ignore this possibility here.} we focus on identifying treatment effect parameters for this set of groups.  

Following existing work on identifying treatment effect parameters under staggered treatment adoption (\citet{callaway-santanna-2021,callaway-li-2021b}), we target identifying group-time average treatment effects
\begin{align*}
    ATT(g,t) = \E[Y_t(g) - Y_t(0) | G=g]
\end{align*}
which is the average effect of participating in the treatment among group $g$ at time period $t$.  Similar to the arguments above, it is straightforward to show that, in periods $t \geq g$ (i.e., post-treatment periods),  
\begin{align} \label{eqn:attgt}
    ATT(g,t) = \E[Y_t - Y_{g-\tau - 2} | G=g] - \Big( \E[X'|G=g] \beta^*_{g,t} + F^*_{g,t} \E[Y_{g-\tau-1} - Y_{g-\tau-2} | G=g]\Big) 
\end{align}
where we define $F^*_{g,t} := \big(F_t-F_{g-\tau-2}\big) \Big/ \big(F_{g-\tau-1} - F_{g-\tau-2}\big)$, $\beta^*_{g,t} := \big(\beta_t - \beta_{g-\tau-2} - F^*_{g,t}(\beta_{g-\tau-1} - \beta_{g-\tau-2}) \big)$.  This suggests that, as earlier, identifying $ATT(g,t)$ comes down to identifying the parameters $\beta^*_{g,t}$ and $F^*_{g,t}$.  We next provide a proposition showing that $\beta^*_{g,t}$ and $F^*_{g,t}$ are identified in post-treatment time periods in a staggered treatment setup with possible anticipation effects.  Before we do that, we define $Z^{stag}_t := \big((1-D_\mathcal{T}) + \indicator{t+\tau \leq \mathcal{T}}(1-D_{t+\tau})D_{\mathcal{T}}\big) Z$ (which is a $K \times 1$ vector that picks up covariates among individuals that (i) are not treated in the current period and are far enough away from participating in the treatment that they are not experiencing anticipation effects, or (ii) that never participate in the treatment), $X^{stag}_g := \big(X', (Y_{g-\tau-1} - Y_{g - \tau - 2})\big)$ (which is a row vector with $K_X + 1$ elements), and $\Delta Y^{post,g-\tau-2} := (Y_t - Y_{g-\tau-2})$.

\begin{proposition} \label{prop:attgt} For any $g \in \mathcal{G}_{\tau}$ and time periods such that $t \geq g$ (i.e., post-treatment time periods for group $g$), and under  \Cref{ass:model,ass:sampling,ass:selection,ass:special-cov,ass:staggered,ass:anticipation} and under the additional conditions that, (i) $\mathcal{T} \geq 3 + \tau$, (ii) $F_{g-\tau-1} \neq F_{g-\tau-2}$, and (iii) that the matrix $\E[Z^{stag}_t {X^{stag}_g}]$ has full rank, and for any $K \times K$ positive definite weighting matrix $\mathbf{W}$,
    \begin{align*}
        \begin{pmatrix} \beta_{g,t}^* \\ F_{g,t}^* \end{pmatrix} = (\E[Z^{stag}_t {X^{stag}_g}]' \mathbf{W} \E[Z^{stag}_t {X^{stag}_g}])^{-1} \E[Z^{stag}_t {X^{stag}_g}]' \mathbf{W} \E[Z^{stag}_t \Delta Y^{post,g-\tau-2}] 
    \end{align*}
    Moreover, for all $g \in \mathcal{G}_\tau$ and time periods such that $t \geq g$, $ATT(g,t)$ is identified and is equal to
    \begin{align*}
        ATT(g,t) = \E[Y_t - Y_{g-\tau - 2} | G=g] - \Big( \E[X'|G=g] \beta^*_{g,t} + F^*_{g,t} \E[Y_{g-\tau-1} - Y_{g-\tau-2} | G=g]\Big)
    \end{align*}
\end{proposition}

The proof of \Cref{prop:attgt} is provided in \Cref{app:additional-proofs} in the Supplementary Appendix.  Given identification of $ATT(g,t)$, in many applications it is common to aggregate these into lower-dimensional parameters of interests.  One popular option is 
\begin{align*}
    ATT^O = \sum_{g \in \mathcal{G}_\tau}  ATT^G(g) \P(G=g | G \in \mathcal{G}_\tau)  
\end{align*}
where $ATT^G(g) := (\mathcal{T} - g + 1)^{-1} \sum_{t=g}^{\mathcal{T}} ATT(g,t)$ which is the average effect of participating in the treatment for group $g$ across its post-treatment time periods.  Another popular option is 
\begin{align*}
    ATT^{ES}(e) = \sum_{g \in \mathcal{G}_\tau} \indicator{g + e \leq \mathcal{T}} \P(G=g|G+e \leq \mathcal{T}, G \in \mathcal{G}_\tau) ATT(g,g+e)
\end{align*}
so that $ATT^O$ corresponds to the average treatment effect across all individuals that ever participate in the treatment (averaged across all their post-treatment time periods), and $ATT^{ES}(e)$ provides the average effect of participating in the treatment by length of exposure to the treatment (ES stands for event study).  See \citet{callaway-santanna-2021} for more aggregations of group-time average treatment effects that could be of interest in particular applications.

\begin{remark}\label{rem:alternative-treatment-regimes}
    This section has considered the setup with staggered treatment adoption.  This framework is common in the literature, but it is possible to allow for more general treatment regimes (for example, allowing for individuals to move into and out of treatment) by defining groups by their entire path of participating or not participating in the treatment.  In most cases, the main data requirement for identifying an $ATT(g,t)$-type parameter with complicated treatment regimes is access to a group of individuals who have not been affected by the treatment up to period $t$.  That being said, more general treatment regimes may run into practical problems (essentially a curse of dimensionality) in applications with a moderate number of observations as the number of observations per treatment path may become small.  See \citet{chaisemartin-dhaultfoeuille-2021a} on difference-in-differences with general treatment regimes.  It would seem possible to adapt those sorts of arguments to our framework.
\end{remark}

\section{Monte Carlo Simulations} \label{sec:mc}

In this section, we provide Monte Carlo simulations illustrating the finite sample properties of our estimation and inference procedure.  In particular, we compare the performance of our approach to difference in differences and linear trends models.  In the Supplementary Appendix, we provide additional Monte Carlo simulations related to (i) weak instruments, (ii) including multiple interactive fixed effects, and (iii) procedures to select the number of interactive fixed effects to include in the model for untreated potential outcomes.

We consider models of the form
\begin{align*}
    Y_{it}(0) = \theta_t + \xi_i + \lambda_iF_t + \alpha W_i + U_{it}
\end{align*}
We consider the case where (i) there are five time periods, (ii) no individual is treated until the fifth time period and there is no effect of participating in the treatment, (iii) $p := \P(D=1) = 0.5$, (iv) $W|D \sim N(0,1)$, (v) $\xi|D,W \sim N(D, 0.1)$ (i.e., $\xi$ is normally distributed with mean $D$ and variance 0.1),  (vi) $\alpha=1$, (vii) we set $\lambda_i = 1 + 2D_i + \rho W_i + \epsilon_i$ where $\epsilon|D,W \sim N(0,0.1)$ so that the mean of the interactive fixed effect $\lambda$ varies across groups and $\rho$ governs how strong of an instrument $W$ is for $\lambda$ (for now, we set $\rho=1$), (viii) $\theta_t = 0.1 (t - 1)$, $U_t|D,W,\xi,\lambda \sim N(0,0.1)$ for all time periods, and (ix) $F_t = t$ for $t=1,\ldots,4$ and we vary $F_5$ among $(4,5,4.5,8)$ across designs.  When $F_5=4$, DID is correctly specified but linear trends is not correctly specified; when $F_5=5$, linear trends is correctly specified but DID is not; setting $F_5=4.5$ implies that neither parallel trends nor linear trends holds as the trend is in between these two cases; when $F_5=8$, the assumptions underlying DID and linear trends are both severely violated.  Finally, we vary $n$ among 100, 500, and 1000.

\begin{table}[t]
\caption{Monte Carlo Simulations Varying Coefficient on IFE}
\label{tab:mc1}
\centering
{ \footnotesize
\begin{tabular}[t]{lrrrrrrrrrrrr}
\toprule
\multicolumn{1}{c}{ } & \multicolumn{3}{c}{Bias} & \multicolumn{3}{c}{RMSE} & \multicolumn{3}{c}{MAD} & \multicolumn{3}{c}{Rej. Prob.} \\
\cmidrule(l{3pt}r{3pt}){2-4} \cmidrule(l{3pt}r{3pt}){5-7} \cmidrule(l{3pt}r{3pt}){8-10} \cmidrule(l{3pt}r{3pt}){11-13}
  & IFE & DID & LT & IFE & DID & LT & IFE & DID & LT & IFE & DID & LT\\
\midrule
\addlinespace[0.3em]
\multicolumn{13}{l}{\textbf{$F_5=4$}}\\
\hspace{1em}$n=100$ & -0.008 & 0.001 & -2.001 & 0.161 & 0.088 & 2.018 & 0.111 & 0.057 & 1.999 & 0.058 & 0.051 & 1.000\\
\hspace{1em}$n=500$ & 0.000 & 0.002 & -1.998 & 0.074 & 0.039 & 2.002 & 0.049 & 0.026 & 1.997 & 0.070 & 0.043 & 1.000\\
\hspace{1em}$n=1000$ & -0.003 & -0.001 & -2.001 & 0.048 & 0.028 & 2.003 & 0.034 & 0.019 & 2.000 & 0.054 & 0.044 & 1.000\\
\addlinespace[0.3em]
\multicolumn{13}{l}{\textbf{$F_5=5$}}\\
\hspace{1em}$n=100$ & -0.006 & 2.008 & 0.006 & 0.290 & 2.022 & 0.159 & 0.172 & 1.999 & 0.101 & 0.073 & 1.000 & 0.062\\
\hspace{1em}$n=500$ & 0.003 & 2.004 & 0.000 & 0.125 & 2.007 & 0.071 & 0.087 & 2.004 & 0.047 & 0.058 & 1.000 & 0.062\\
\hspace{1em}$n=1000$ & -0.003 & 2.002 & -0.001 & 0.083 & 2.003 & 0.050 & 0.059 & 2.002 & 0.034 & 0.043 & 1.000 & 0.052\\
\addlinespace[0.3em]
\multicolumn{13}{l}{\textbf{$F_5=4.5$}}\\
\hspace{1em}$n=100$ & -0.008 & 0.996 & -1.007 & 0.208 & 1.005 & 1.024 & 0.133 & 0.994 & 1.006 & 0.050 & 1.000 & 1.000\\
\hspace{1em}$n=500$ & 0.003 & 1.000 & -0.993 & 0.095 & 1.002 & 0.997 & 0.062 & 1.003 & 0.995 & 0.066 & 1.000 & 1.000\\
\hspace{1em}$n=1000$ & 0.000 & 1.002 & -0.997 & 0.066 & 1.003 & 0.999 & 0.045 & 1.002 & 0.997 & 0.055 & 1.000 & 1.000\\
\addlinespace[0.3em]
\multicolumn{13}{l}{\textbf{$F_5=8$}}\\
\hspace{1em}$n=100$ & -0.037 & 7.966 & 5.972 & 0.757 & 8.011 & 6.008 & 0.507 & 7.960 & 5.972 & 0.064 & 1.000 & 1.000\\
\hspace{1em}$n=500$ & -0.008 & 7.999 & 5.998 & 0.334 & 8.008 & 6.005 & 0.227 & 8.003 & 5.995 & 0.053 & 1.000 & 1.000\\
\hspace{1em}$n=1000$ & 0.008 & 7.986 & 5.990 & 0.226 & 7.990 & 5.994 & 0.156 & 7.985 & 5.996 & 0.053 & 1.000 & 1.000\\
\bottomrule
\end{tabular}}
\begin{justify}
{\footnotesize \textit{Notes:} The columns labeled ``IFE'' use the interactive fixed effect approach introduced in the paper; the columns labeled ``DID'' provide results using a difference in differences approach; the columns labeled ``LT'' provide results using a linear trends approach.  Columns labeled ``Bias'' simulate the bias of each approach, ``RMSE'' simulates the root mean squared error, and ``MAD'' for the median absolute deviation.  Columns labeled ``Rej. Prob.'' contain the rejection probabilities for a test of the null hypothesis that $ATT_5=0$ (which is true here) at the 5\% significance level.  The rows vary $F_5$ and $n$ as discussed in the text.  Results come from 1000 Monte Carlo simulations.}
\end{justify}
\end{table}

The results from the first set of simulations are provide in \Cref{tab:mc1}.  These results are broadly in line with our earlier discussions.  IFE generally performs well across each different value of $F_5$, and all estimators tend to perform somewhat better with larger sample sizes.  In the case where DID is correctly specified (i.e., when $F_5=4$), the DID estimator performs somewhat better in terms of root mean squared error and median absolute deviation than IFE though, as expected, linear trends does not perform well in this case.  Similarly, in the case where linear trends is correctly specified (i.e., when $F_5=5$), linear trends performs somewhat better in terms of root mean squared error and median absolute deviation than IFE though both do substantially better than DID.  In the intermediate case where $F_5=4.5$, IFE substantially outperforms both DID and linear trends for all sample sizes.  The last case that we consider is when $F_5=8$, so that DID and linear trends are substantially misspecified.  In this case, IFE continues to perform well while DID and linear trends perform poorly.

The inference procedure for IFE also works well across simulation setups.  Across all sample sizes and values of $F_5$, our IFE approach rejects the null of $ATT_5=0$ (which is true) at close to its nominal level.  For DID and linear trends, inference works well in cases where they are correctly specified, but (as expected) there are severe distortions (particularly, over-rejecting) when they are not correctly specified.

\section{Application} \label{sec:job-displacement}

In this section, we consider an application on the effect of early-career job displacement on earnings.  Much of the literature on job displacement has been interested in the dynamics of the effects of job displacement (see, for example, \citet{jacobson-lalonde-sullivan-1993} as well as many subsequent papers).  Difference in differences has been a standard approach in this literature with the idea being that displaced workers may differ from non-displaced workers along dimensions, for example latent ``skill'', that are unobserved by the researcher.

One potential issue with this approach is that the return to latent skill could vary over time (e.g., with the business cycle) as well as vary with experience.  If skill were observed, it would be natural to include it in the model for untreated (non-displaced) potential outcomes while allowing its coefficient to vary over time.  Although standard difference in differences approaches cannot handle this sort of issue, this setup mirrors the approach discussed in the current paper exactly.  In addition, like the difference in differences approach, our setup allows for other unobserved, time invariant variables to affect an individual's earnings in the absence of displacement and be distributed differently across displaced and non-displaced workers as long as their effect is time invariant as well.

\subsection{Data and Setup}

\newcolumntype{.}{D{.}{.}{-1}}
\ctable[caption={Summary Statistics}, label=tab:ss, pos=t,]{lD{.}{.}{-1}D{.}{.}{-1}D{.}{.}{-1}D{.}{.}{-1}}{\tnote[]{\textit{Notes:} The column ``Displaced'' aggregates summary statistics for workers that were displaced from their jobs between 1989 and 1993.  Reported earnings are yearly and reported in thousands of dollars.  The column ``Diff'' reports the difference between displaced and non-displaced workers.  The column ``p-val on diff'' reports the p-value of a test for the equality of each variable for displaced and non-displaced workers.

  \textit{Sources:}  1979 National Longitudinal Study of Youth}}{\FL
\multicolumn{1}{l}{}&\multicolumn{1}{c}{Displaced}&\multicolumn{1}{c}{Non-Displaced}&\multicolumn{1}{c}{Diff}&\multicolumn{1}{c}{p-val on diff}\ML
{\bfseries Path of Earnings}&&&&\NN
~~Earnings 1983&~8.29&~9.21&~-0.91&0.02\NN
~~Earnings 1985&11.67&13.08&~-1.40&0.00\NN
~~Earnings 1987&14.82&16.69&~-1.88&0.00\NN
~~Earnings 1989&16.59&20.37&~-3.78&0.00\NN
~~Earnings 1991&17.14&23.28&~-6.14&0.00\NN
~~Earnings 1993&18.42&25.60&~-7.19&0.00\ML
{\bfseries Covariates}&&&&\NN
~~Less HS&~0.13&~0.07&~~0.06&0.00\NN
~~HS&~0.64&~0.59&~~0.05&0.04\NN
~~College&~0.23&~0.34&~-0.11&0.00\NN
~~Hispanic&~0.23&~0.16&~~0.07&0.00\NN
~~Black&~0.32&~0.23&~~0.09&0.00\NN
~~White&~0.45&~0.61&~-0.16&0.00\NN
~~Male&~0.57&~0.45&~~0.12&0.00\NN
~~AFQT&39.92&51.36&-11.44&0.00\NN[10pt]
~~N & 416 & 2434 & & \LL
}

We use data from the 1979 National Longitudinal Study of Youth (NLSY).  The NLSY is an ongoing panel data survey of U.S. residents that were between 14 and 22 when the survey was started in 1979.  Following standard approaches in the job displacement literature, we define workers as being displaced if they report no longer being at the same job as in the previous survey and the reason that they left their job is (i) layoff, (ii) plant closed, or (iii) company, office, or workplace closed.  Importantly, this does not include individuals who were fired from their job or quit.  

The main outcome that we consider is yearly earnings, and we use bi-annual earnings data from 1983-1993.  Our treated groups come from workers who report being displaced from their job in the 1989, 1991, or 1993 waves of the survey.  We also include workers that were not displaced in any period through 1993, but drop all individuals who were displaced in any period before 1989.  To be included in the sample, we additionally require that individuals had positive earnings in both 1983 and 1985.  This results in a total sample size of 2,850 individuals.  For covariates, we observe workers' educational attainment, race, and gender --- all of these characteristics are commonly observed in micro-level data used in labor economics.  In addition to these covariates, one of the interesting variables in the NLSY is an individual's score on the Armed Forces Qualification Test (AFQT).  Researchers using the NLSY have frequently used an individual's AFQT score as a proxy for their unobserved skill/ability, most commonly including AFQT score directly as an additional covariate.  

Summary statistics are provided in \Cref{tab:ss}.  To start with, it is worth pointing out a few suggestive patterns in the summary statistics.   First, notice that in 1983 (before any workers in our sample have been displaced), non-displaced workers already have higher earnings than displaced workers.  This suggests that making comparisons based on only the level of earnings over time for displaced and non-displaced workers is likely to lead to misleadingly large estimates of the effect of job displacement.  Second, in 1985 (still before any workers in our sample have been displaced), the gap between earnings of displaced and non-displaced workers increases.  This further suggests that, at least between 1983 and 1985, the trend in earnings over time was not the same for displaced and non-displaced workers which implies that a difference in differences approach to estimating the effect of job displacement would not have performed well (and would have been likely to over-estimate the effect of job displacement).

In addition, the gap between the earnings of displaced and non-displaced workers also tends to continue to widen over time.  There are also noticeable differences in terms of other covariates.  Relative to displaced workers, non-displaced workers tend to be more educated, more likely to be white and less likely to be black or Hispanic, more likely to be female, and have higher AFQT scores.

Determining which covariate(s) have time invariant effects on untreated potential outcomes is an important decision in our framework.  We take two approaches here.  First, we use AFQT score in this role (that is, in \Cref{eqn:3}, we treat unobserved skill as $\lambda$ and AFQT score as $W$.)   The case for using AFQT in this role is strong.  Since AFQT is often seen as a proxy for skill, it is probably reasonable to think that AFQT meets the stronger requirement that it does not affect earnings directly at all as long as ``skill'' is in the model.  For our approach, we only require that the \textit{path} of earnings in the absence of job displacement does not depend on AFQT as long as skill is in the model.  

Having access to a proxy variable like AFQT is not typical in applications though.  For this reason, we also provide results where we use educational attainment as the covariate whose effect on untreated potential outcomes does not change over time (once skill is included in the model).  The case that education does not have an effect on the path of untreated potential outcomes after including skill in the model is not as strong as it was for AFQT score; however, because there is very limited variation in educational attainment over time for adults, it is often not included in panel data regressions estimating effects of job displacement.  This at least tentatively suggests that educational attainment is a plausible candidate for a covariate with time invariant effects.  And, in particular, these arguments could hold if education affects an individual's first job or occupation but, after that, it is skill that affects the path of earnings. We stress that our results that use AFQT score are likely more credible than the ones that use education, but the reasons that we also consider this case are (i) educational variables are widely available, (ii) our pre-tests about whether a variable affects the path of untreated potential outcomes are relevant in this case, and (iii) it is interesting to compare these results to the ones that come from using AFQT.  

There are a few remaining practical questions that we are able to address in the application.  First, there is variation in treatment timing; i.e., the timing of displacement is not constant across all individuals.  Second, a common finding in the job displacement literature (see, for example, \citet{jacobson-lalonde-sullivan-1993} and much subsequent work) is that earnings ``dip'' before job displacement actually takes place (this phenomenon is typically explained by firms struggling before there is a mass layoff; this can result in reduced hours for their employees and/or slow wage growth in periods before workers are actually laid off).  We can address both of these issues along the lines discussed in  \Cref{sec:staggered}.  In particular, we allow for anticipation by setting $\tau=2$ (i.e., two years of anticipation) and then compute group-time average treatment effects  as in \Cref{eqn:attgt}.  
We report our results below in event study plots.  In pre-treatment periods, these results can be used to ``pre-test'' the identifying assumptions.  In our case, where we are interested in the relative performance of several different approaches to identifying and estimating the effects of job displacement, it is particularly useful to compare the performance of each approach in pre-treatment time periods.  In post-treatment time periods, these results can be used to study the dynamics of the effect of job displacement.

\subsection{Results}

{ \setlength{\tabcolsep}{10pt}
\begin{table}[t!] 
\centering
\caption{Estimates of Interactive Fixed Effects Model for Untreated Potential Outcomes}
\label{tab:model}
\begin{table}[H]
\centering

(a) Pre-Treatment Periods
\begin{tabular}[t]{lcccccc}
\toprule
  & g:89,t:87 & g:91,t:87 & g:91,t:89 & g:93,t:87 & g:93,t:89 & g:93,t:91\\
\midrule
(Intercept) & \num{-0.78} & \num{-0.73} & \num{-0.18} & \num{-0.82} & \num{0.02} & \num{-0.69}\\
 & (\num{0.85}) & (\num{0.86}) & (\num{0.85}) & (\num{0.91}) & (\num{0.85}) & (\num{1.02})\\
IFE & \num{2.14}* & \num{2.12}* & \num{2.06}* & \num{2.15}* & \num{2.01}* & \num{1.98}*\\
 & (\num{0.21}) & (\num{0.21}) & (\num{0.22}) & (\num{0.23}) & (\num{0.22}) & (\num{0.26})\\
\midrule
N & \num{2721} & \num{2567} & \num{2567} & \num{2434} & \num{2434} & \num{2434}\\
Weak IV F-stat & \num{70.05} & \num{67.27} & \num{59.30} & \num{62.18} & \num{56.75} & \num{41.28}\\
\bottomrule
\multicolumn{7}{l}{\rule{0pt}{1em}* p $<$ 0.05}\\
\end{tabular}

\bigskip

\bigskip

(b) Post-Treatment Periods

\begin{tabular}[t]{lcccccc}
\toprule
  & g:89,t:89 & g:89,t:91 & g:89,t:93 & g:91,t:91 & g:91,t:93 & g:93,t:93\\
\midrule
(Intercept) & \num{-1.68} & \num{-3.11} & \num{-3.24} & \num{-0.66} & \num{-0.34} & \num{-0.38}\\
 & (\num{1.45}) & (\num{2.10}) & (\num{2.43}) & (\num{1.41}) & (\num{1.66}) & (\num{1.35})\\
IFE & \num{3.31}* & \num{4.44}* & \num{5.08}* & \num{3.00}* & \num{3.56}* & \num{2.53}*\\
 & (\num{0.36}) & (\num{0.52}) & (\num{0.60}) & (\num{0.37}) & (\num{0.43}) & (\num{0.35})\\
\midrule
N & \num{2567} & \num{2434} & \num{2434} & \num{2434} & \num{2434} & \num{2434}\\
Weak IV F-stat & \num{67.27} & \num{62.18} & \num{62.18} & \num{56.75} & \num{56.75} & \num{41.28}\\
\bottomrule
\multicolumn{7}{l}{\rule{0pt}{1em}* p $<$ 0.05}\\
\end{tabular}

\end{table}
\begin{justify}
{\small \textit{Notes:}  The table contains estimates of the interactive fixed effects model for untreated potential outcomes using AFQT as the covariate whose effect on untreated potential outcomes does not change over time and without including other covariates. Columns correspond to estimates for particular groups and particular time periods.  For example, the first column in panel (a), which is labeled ``g:89,t:87'' is for the group of workers who were displaced in 1989 in the year 1987.  The rows labeled ``IFE'' report the estimated value of $F_{g,t}^*$.  Rows labeled ``Weak IV F-stat'' report the F-statistic from the first stage regression of the endogenous regressors on AFQT score.  Some F-statistics are the same because the first stage is the same across some groups/time periods.}
\end{justify}
\end{table}
}

To start with, we estimate the interactive fixed effects models for untreated potential outcomes.  In our setup with staggered treatment adoption, we estimate the model separately for each group and time period --- this results in six models to estimate in pre-treatment time periods and six models to estimate in post-treatment time periods.  To estimate each model, we use the group of workers that are not displaced from their job in any period as well as groups of workers who eventually become displaced but are ``far enough'' away from displacement that they are not yet affected (including anticipation effects).\footnote{To give a more concrete example, for the group of workers that are displaced in 1989, in 1987 their comparison group includes never-displaced workers and workers displaced in 1991 and 1993; in 1989 their comparison group includes never-displaced workers and workers displaced in 1993; and in 1991 and 1993, their comparison group includes only never-displaced workers.}  These estimates are provided in \Cref{tab:model}.  Interpreting our estimates of $F_{g,t}^*$, the coefficient on the change in outcomes in pre-treatment periods, as being due to changes in the return to skill over time, our estimates indicate that the return to skill was tending to increase over the time periods that we consider (in the absence of job displacement).  These effects appear relatively large and are strongly statistically significant.  \Cref{tab:model} also provides an F-test for weak instruments in the first step estimates and strongly rejects the null of the instrument being weak (more details about the first stage regression are provided in \Cref{tab:afqt-first-stage} in the Supplementary Appendix).

\begin{figure}[t!] 
    \centering
    \caption{Effect of Job Displacement on Earnings}
    \label{fig:ife_no_covs}
    \begin{subfigure}[b]{.49\textwidth}
    \includegraphics[width=\textwidth]{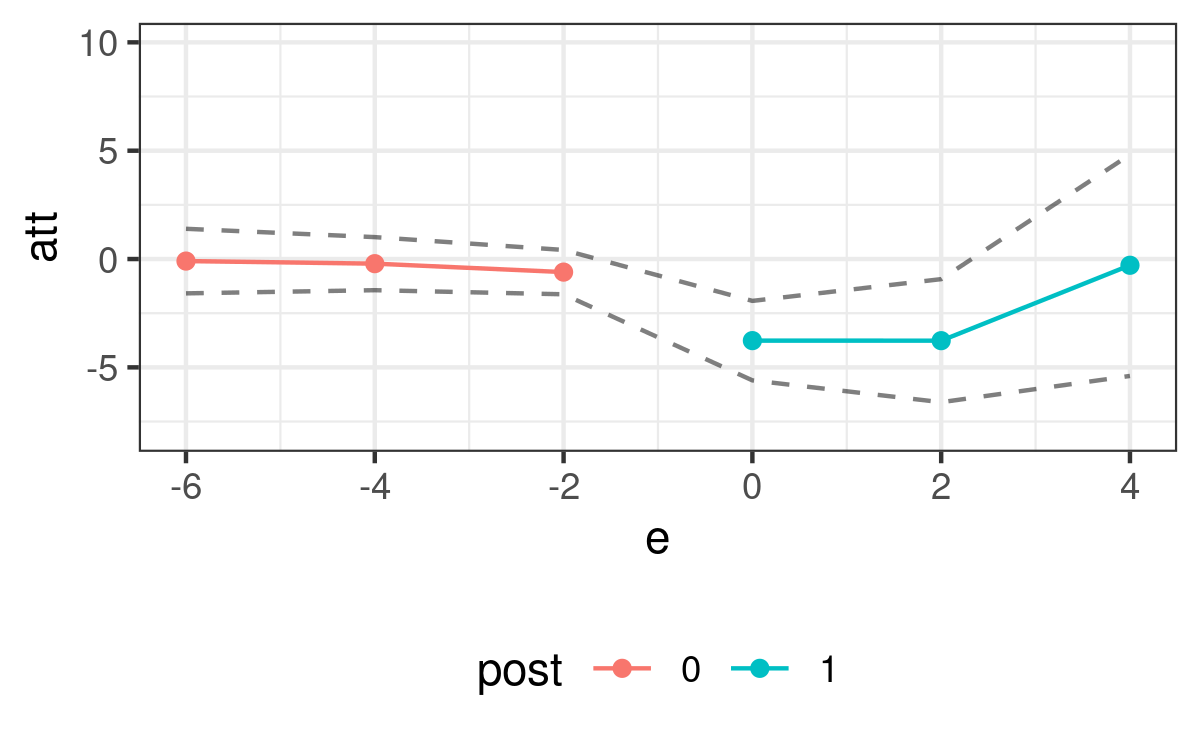}
    \caption{IFE}
    \end{subfigure}
    \begin{subfigure}[b]{.49\textwidth}
    \includegraphics[width=\textwidth]{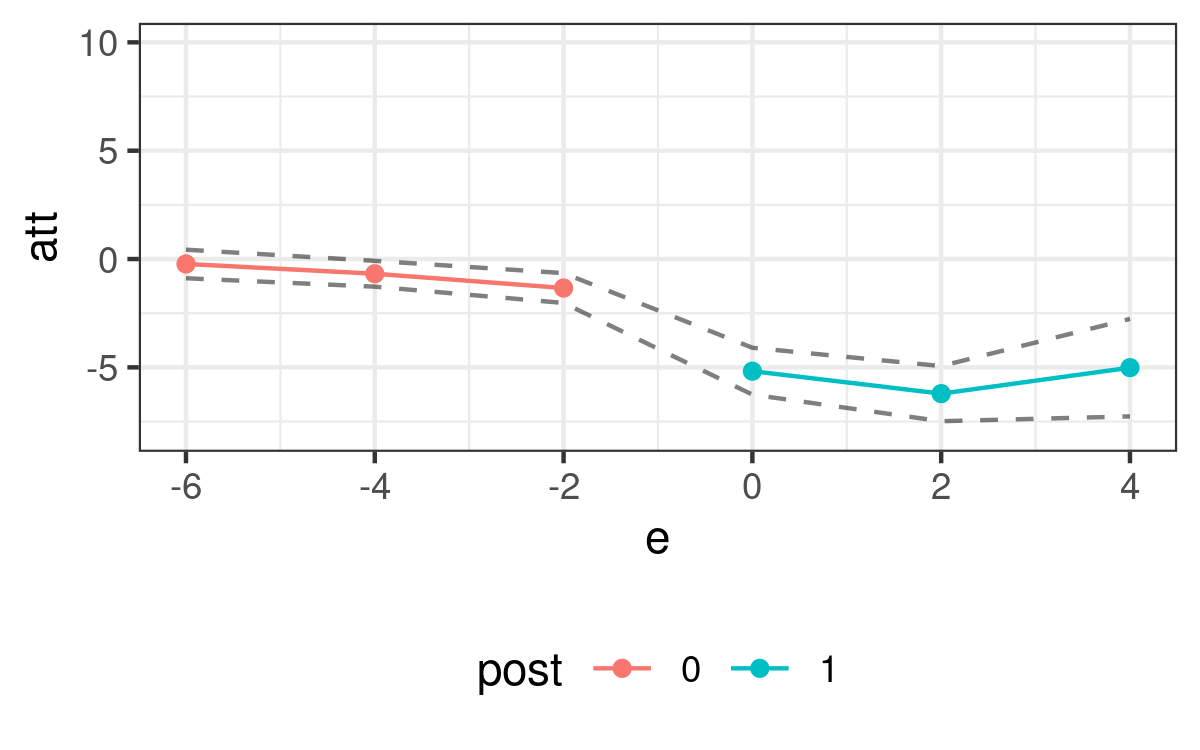}
    \caption{DID}
    \end{subfigure}
    \begin{subfigure}[b]{.49\textwidth}
    \includegraphics[width=\textwidth]{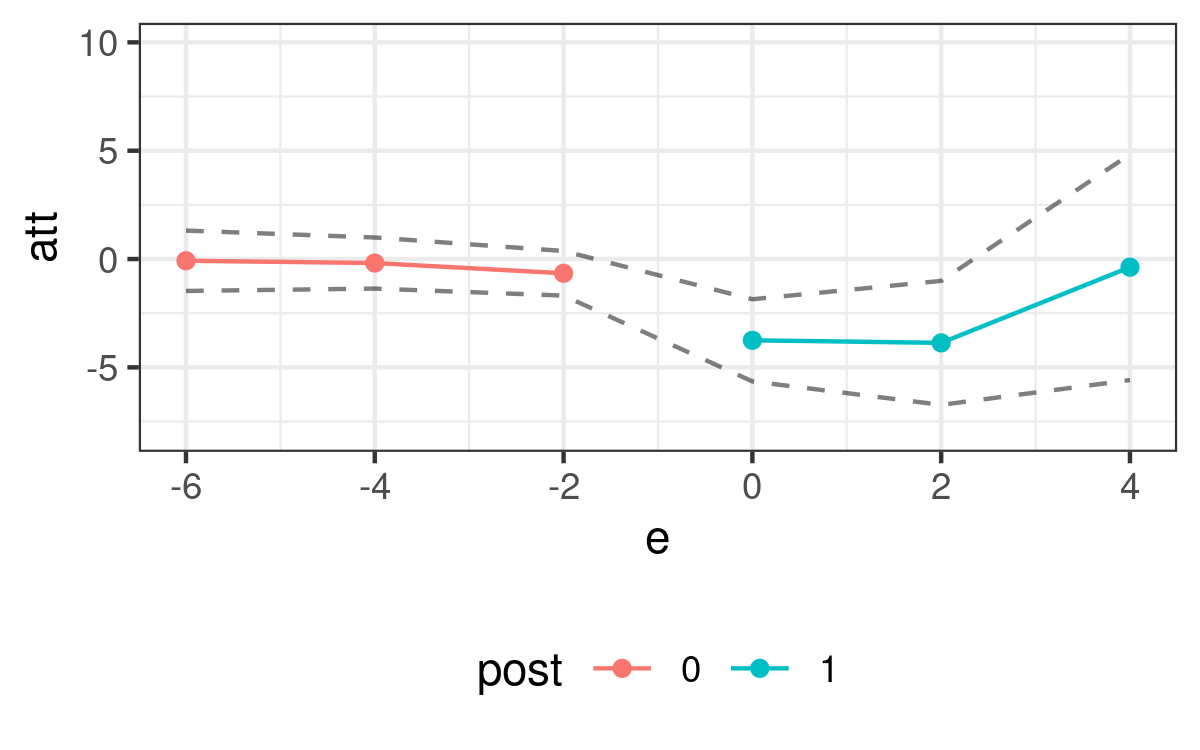}
    \caption{Linear Trends}
    \end{subfigure}
    
    \begin{justify}
    { \footnotesize \textit{Notes:} The figure contains estimates of the effect of job displacement on yearly earnings (which are in thousands of dollars) by length of exposure to job displacement.  Negative values of $e$ correspond to the number of years before job displacement occurred, $e=0$ in the year that job displacement occurred, and positive values of $e$ indicate how long ago an individual was displaced from their job.  Panel (a) contains estimates from the interactive fixed effects approach proposed in the current paper using AFQT as the covariate with time invariant effects on untreated potential outcomes.  Panels (b) and (c) contain analogous estimates but using difference in differences and linear trends models, respectively.  Each panel provides 90\% confidence intervals computed using the multiplier bootstrap. 
    }
    \end{justify}
\end{figure}

Next, we provide our main results on estimating the effect of job displacement.  The first set of main results are provided in \Cref{fig:ife_no_covs}.  To begin with, consider the DID estimates in panel (b).  Two years before job displacement takes place, the DID estimate of the effect of job displacement is negative (slightly more than \$1,300 lower earnings due to job displacement) and statistically significant (p-value: 0.001).  This is perhaps explained by a pre-displacement dip in earnings discussed above.  However, even four years before job displacement, the DID estimates indicate that displaced workers have almost \$700 lower earnings due to job displacement (p-value: 0.061).  Together, these are suggestive that the trend in earnings, even in the absence of job displacement, is not the same for displaced and non-displaced workers.  It is interesting to compare the DID estimates in pre-treatment periods to the ones that come from using our approach, which are provided in panel (a).  There we estimate notably smaller effects in pre-displacement periods (about \$200 lower earnings four years before displacement and about \$600 lower two years before displacement; recall that the latter estimates may be influenced by a pre-displacement earnings dip due to job displacement); neither of these estimates are statistically different from zero.  

Whether using DID or our IFE approach, the estimated effects of job displacement are relatively similar on impact and two years following job displacement --- about \$3,700 lower earnings on average due to job displacement at both lengths of exposure using IFE estimates and about \$5,200 on impact and \$6,200 two years following job displacement for DID.  However, there are notable differences in the estimates four years following job displacement.  The IFE estimates are small (earnings losses of \$300 on average and not statistically different from zero) while the DID estimates are large and similar in magnitude to the estimates in earlier periods (about \$5000 lower earnings on average due to job displacement); the standard errors are relatively large but the difference between the two estimates is marginally statistically significant (p-value: 0.074).  While not conclusive, these differences are at least quite interesting in the context of the job displacement literature because, like our DID estimates, the DID approaches that are common in the job displacement literature often deliver large, negative estimates of the long-term effect of job displacement (e.g., \citet{jacobson-lalonde-sullivan-1993,wachter-song-manchester-2009}).  These long-lasting effects  are somewhat challenging to rationalize (that said, there are some plausible explanations such as displaced workers losing access to firm-specific wage premiums that are very difficult to recover).  An alternative explanation that is compatible with our results is that displaced and non-displaced workers have different skill distributions that cause violations of parallel trends (both in pre-treatment and post-treatment time periods) and that this leads to over-estimates of the long-term effects of job displacement (i.e., even in the absence of job displacement, the trend in earnings for displaced workers would have been less than the trend in earnings for non-displaced workers).  We also note that estimates from a linear trends model for untreated potential outcomes are very similar to estimates using our IFE approach.

\begin{figure}[t!]
    \centering
    \caption{Effect of Job Displacement on Earnings without AFQT}
    \label{fig:educ}
    \begin{subfigure}[b]{.49\textwidth}
    \includegraphics[width=\textwidth]{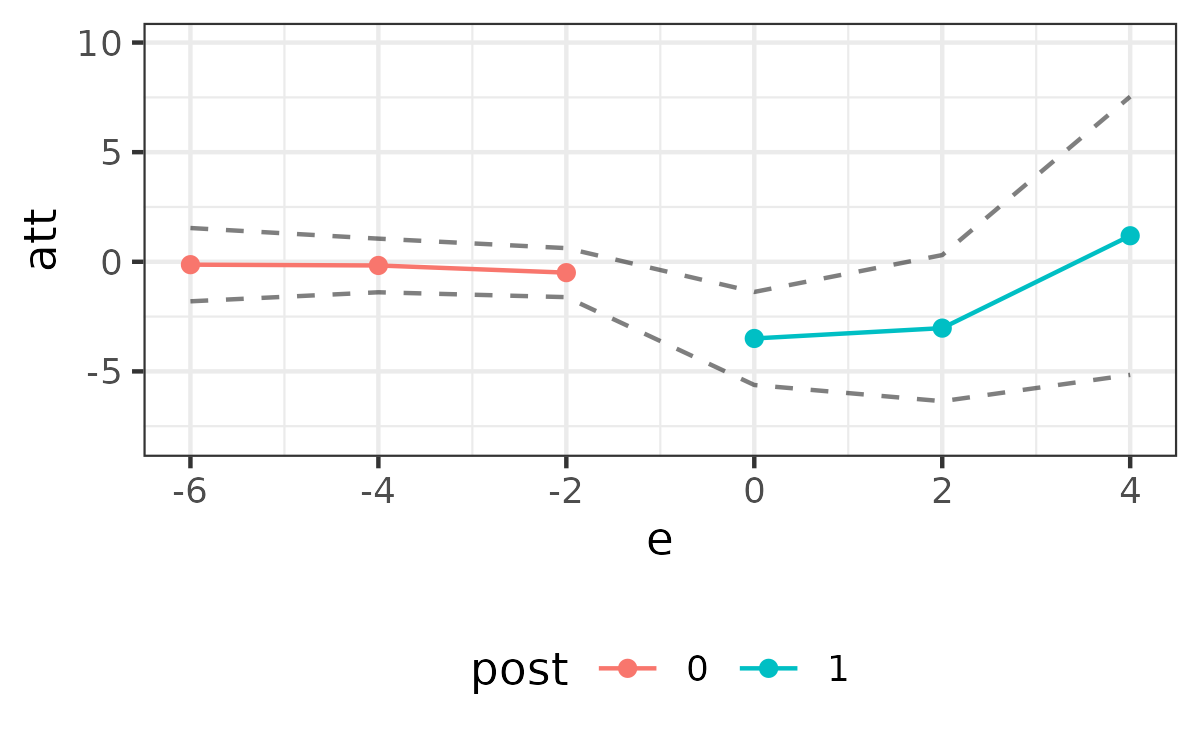}
    \caption{IFE, Time invariant education}
    \end{subfigure}
    \begin{subfigure}[b]{.49\textwidth}
    \includegraphics[width=\textwidth]{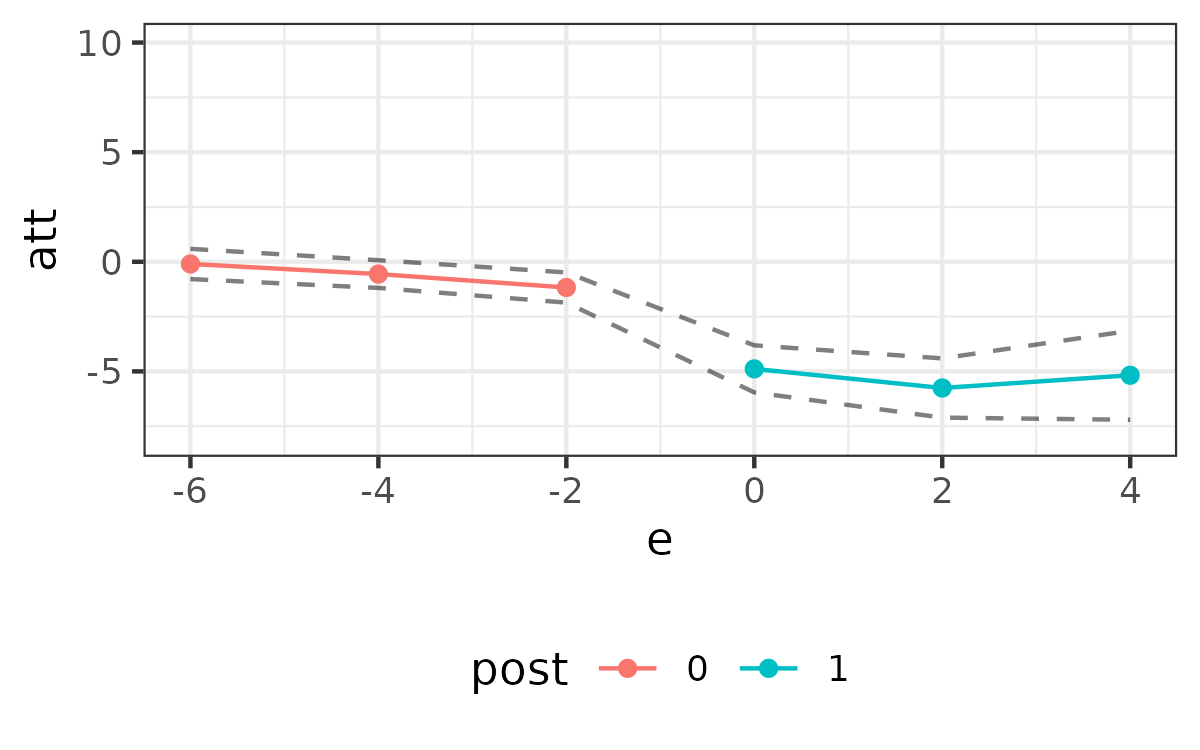}
    \caption{DID, Education-specific trends}
    \end{subfigure}
    
    \begin{justify}
    { \footnotesize \textit{Notes:} The figure contains estimates of the effect of job displacement on yearly earnings (which are in thousands of dollars) by length of exposure to job displacement.  Negative values of $e$ correspond to the number of years before job displacement occurred, $e=0$ in the year that job displacement occurred, and positive values of $e$ indicate how long ago an individual was displaced from their job.  Panel (a) contains estimates from the interactive fixed effects approach proposed in the current paper using educational attainment as the covariate with time invariant effects on untreated potential outcomes.  Panel (b) includes education as a covariate (which can have time varying effects) but does not include an interactive fixed effects term.  Each panel provides 90\% confidence intervals computed using the multiplier bootstrap. 
    }
    \end{justify}
\end{figure}

Next, we move to considering the thought-experiment where the proxy variable AFQT is not available.  In this case, we use educational attainment as the covariate whose effect does not change over time.  These results are provided in \Cref{fig:educ}, and first-step estimates of the model for untreated potential outcomes are provided in \Cref{tab:ife-educ} in the Supplementary Appendix.  These results are quite similar to the previous ones using AFQT as the covariate whose effect does not change over time though the standard errors are somewhat wider.  As before, these results (in panel (a))  indicate a large negative effect of job displacement on impact and two years following job displacement potentially followed by smaller effects four years following job displacement (though noting that standard errors are quite large in this last case especially).  These results are notably different from the DID estimates in panel (b) that use educational attainment as a covariate (i.e., these results do not include an interactive fixed effect in the model for untreated potential outcomes but do include education as a covariate whose effect can change over time).  In this latter case,  the estimates of the effects of job displacement are large and continue to be large four years following job displacement.

In \Cref{app:additional-application-results}, we provide results for some additional cases.  First, when the researcher additionally includes education, race, and gender as covariates while using AFQT score as the covariate with time invariant effects, we find very similar results to the ones reported in \Cref{fig:ife_no_covs}; similarly, when we include race and gender as covariates while using educational attainment as the covariate with time invariant effects, we find very similar results to the ones reported in \Cref{fig:educ} (see \Cref{fig:ife_covs} for these results).  Second, we provide estimates of the model for untreated potential outcomes in pre-treatment time periods when the model is identified without requiring a covariate to have a time invariant effect, i.e., along the lines of \Cref{sec:time-invariant-effects-test}.  We provide these results in \Cref{tab:time-invariant-covariates-test} where we report estimates from using earnings in 1987 as the outcome and using group membership (among those first displaced in 1991, 1993, or the never-displaced group --- we do not use those first displaced in 1989 due to possible anticipation effects) as the extra set of moment conditions to identify the model.  The pre-treatment interactive fixed effects models for untreated potential outcomes estimated using this approach are generally estimated imprecisely.\footnote{Among groups first treated in 1991, 1993, and the never-displaced group, paths of earnings from 1983 to 1985 are relatively similar which leads to a weak instrument problem; interestingly, for the group first-treated in 1989, their path of earnings from 1983 to 1985 is significantly different from the other groups, but we are not able to use that group here because they may be experiencing anticipation effects in 1987.}  That said, there are still some interesting patterns that emerge.  Across specifications, the estimated coefficients on AFQT and educational attainment variables notably shrink in magnitude when interactive fixed effects are included in the model.  These results are generally supportive of (and, at a minimum, do not provide any evidence against) both the identification strategy using AFQT as the covariate whose effect does not vary over time as well as using educational attainment in the same role.   Finally, we consider the case with two interactive fixed effects where we use both AFQT and education as covariates with time invariant effects; see \Cref{fig:multiple-ifes} where we report results separately for cases where race and gender are or are not included as covariates.  These results are also broadly similar to the IFE results presented so far in this section though the standard errors tend to be larger, and we are not able to estimate the event study for $e=-6$ or $e=4$ due to requiring more time periods when there are more interactive fixed effects, as discussed in \Cref{sec:multiple-ife}.

\FloatBarrier

\section{Conclusion}

In this paper, we have developed a new approach to identifying and estimating the ATT when untreated potential outcomes have an interactive fixed effects structure.  Interactive fixed effects models are relevant models in the context of policy evaluation in the case where the effect of some time invariant unobservable can vary over time.  Unlike common approaches such as linear trends models, interactive fixed effects models allow for the time invariant unobservable to enter the model for untreated potential outcomes in an analogous way to observed covariates.

There are a number of attractive features of our approach for use in applied work.  First, our approach generalizes the most common approaches to policy evaluation (difference in differences and linear trends models) when a researcher has access to repeated observations over time.  Second, our approach only requires that the researcher has access to at least three periods of data rather than requiring that the number of time periods goes to infinity as is often a requirement in the literature on interactive fixed effects models.  Third, our approach can be used when a researcher has access to either panel or repeated cross sections data.  Fourth, our approach allows for very general forms of treatment effect heterogeneity and selection into participating in the treatment.  In addition, we discussed how our approach can accommodate multiple interactive fixed effects and accommodate variation in treatment timing and treatment anticipation.  The main requirement for using our approach is access to a time invariant variable whose effect on untreated potential outcomes does not change over time (though the variable can affect the level of the outcome).  It seems likely that this sort of variable will exist in many applications in economics.  A good candidate for this sort of variable is any variable that the researcher would not include in the model because it does not vary over time.

In applications, the three main challenges for using our approach are likely to be (i) weak instrument problems, (ii) if the number of covariates with time invariant effects is less than the number of interactive fixed effects, and (iii) uncertainty about particular covariates actually having time invariant effects.  Another attractive feature of our approach though is that researchers should generally be able to, at a minimum, successfully diagnose each of these issues in applications.  First, well-known weak instrument diagnostics apply immediately to our framework.  Second, our approach inherits many of the most important advantages of DID related to pre-testing the identifying assumptions in periods before the treatment took place.  In particular, the number of interactive fixed effects being greater than the number of covariates with time invariant effects will generally lead to biased (i.e., non-zero) ``effects'' in pre-treatment periods suggesting a failure of the identifying assumptions.  Third, in order to address uncertainty regarding whether or not a particular covariate has a time invariant effect, we developed a ``pre-test'' for this condition as well (and additionally note that, when available, over-identification tests also can be useful here).  Taken together, this suggests that our approach opens up new possibilities for researchers to identify and estimate causal effect parameters when repeated observations are available while cases where our approach will not work well can be detected.

Finally, the computational burden of our procedure is low --- in our application on job displacement, our event study estimates, across all event times and using 1000 multiplier bootstrap iterations, take 4.5 seconds on a 2.80-GHz Intel i5 processor with 8GB of RAM and with no parallel processing.  Code for our approach is available in the \texttt{R} \texttt{ife} package which is available at \url{https://github.com/bcallaway11/ife}.

\singlespacing

\printbibliography

\pagebreak

\appendix

\numberwithin{equation}{section}
\numberwithin{table}{section}
\numberwithin{figure}{section}

\FloatBarrier

\section{Additional Results on Job Displacement} \label{app:additional-application-results}

This section contains additional figures, tables, and results from the application on job displacement in the main text.

\begin{table}[ht]
\centering
\caption{Pre-Treatment Estimates of Model for Untreated Potential Outcomes}
\label{tab:time-invariant-covariates-test}
\begin{tabular}[t]{lcccccccc}
\toprule
\multicolumn{1}{c}{ } & \multicolumn{4}{c}{IFE} & \multicolumn{4}{c}{No IFE} \\
\cmidrule(l{3pt}r{3pt}){2-5} \cmidrule(l{3pt}r{3pt}){6-9}
  & afqt & afqt, covs & educ & educ, covs & afqt  & afqt, covs  & educ  & educ, covs \\
\midrule
(Intercept) & \num{-0.44} & \num{-0.54} & \num{-0.24} & \num{0.27} & \num{3.29}* & \num{4.77}* & \num{4.16}* & \num{5.75}*\\
 & (\num{3.70}) & (\num{7.18}) & (\num{8.22}) & (\num{7.87}) & (\num{0.38}) & (\num{0.76}) & (\num{0.67}) & (\num{0.75})\\
IFE & \num{1.96} & \num{1.68} & \num{1.66} & \num{1.50} &  &  &  & \\
 & (\num{1.93}) & (\num{2.26}) & (\num{3.09}) & (\num{2.14}) &  &  &  & \\
AFQT & \num{0.01} & \num{0.01} &  &  & \num{0.08}* & \num{0.06}* &  & \\
 & (\num{0.08}) & (\num{0.07}) &  &  & (\num{0.01}) & (\num{0.01}) &  & \\
High School &  & \num{0.79} & \num{0.97} & \num{1.19} &  & \num{1.26} & \num{1.91}* & \num{2.67}*\\
 &  & (\num{0.99}) & (\num{1.90}) & (\num{2.22}) &  & (\num{0.73}) & (\num{0.72}) & (\num{0.71})\\
College &  & \num{1.75} & \num{2.21} & \num{2.73} &  & \num{4.39}* & \num{6.59}* & \num{7.25}*\\
 &  & (\num{3.66}) & (\num{8.19}) & (\num{6.50}) &  & (\num{0.85}) & (\num{0.75}) & (\num{0.75})\\
Black &  & \num{0.72} &  & \num{0.54} &  & \num{0.57} &  & \num{-0.11}\\
 &  & (\num{0.66}) &  & (\num{1.09}) &  & (\num{0.60}) &  & (\num{0.60})\\
White &  & \num{0.29} &  & \num{0.39} &  & \num{-0.54} &  & \num{0.25}\\
 &  & (\num{1.25}) &  & (\num{0.55}) &  & (\num{0.53}) &  & (\num{0.53})\\
Female &  & \num{-0.59} &  & \num{-1.02} &  & \num{-4.25}* &  & \num{-4.39}*\\
 &  & (\num{4.93}) &  & (\num{4.83}) &  & (\num{0.37}) &  & (\num{0.37})\\
\midrule
N & \num{2721} & \num{2721} & \num{2721} & \num{2721} & \num{2721} & \num{2721} & \num{2721} & \num{2721}\\
Sargan p-value & \num{0.918} & \num{0.849} & \num{0.834} & \num{0.915} &  &  &  & \\
Weak IV F-stat & \num{0.36} & \num{0.22} & \num{0.11} & \num{0.21} &  &  &  & \\
\bottomrule
\multicolumn{9}{l}{\rule{0pt}{1em}* p $<$ 0.05}\\
\end{tabular}
\begin{justify}
    { \footnotesize \textit{Notes:} The table contains estimates for models of untreated potential outcomes in 1987 using observations for groups first treated in 1991, 1993, and the never-treated group with extra moment conditions coming from group membership.  Each column varies the covariates included in the model.  The first four columns provide estimates that include an interactive fixed effect, and, for comparison, the last four columns provide estimates that do not include an interactive fixed effect.  Rows labeled ``Sargan p-value'' provide a p-value for a Sargan over-identification test (which is available because there are three groups and only one interactive fixed effect); rows labeled ``Weak IV F-stat'' report the F-statistic from the first stage regression of the endogenous regressors on group dummies and exogenous regressors.
    }
\end{justify}
\end{table}

\begin{figure}[t]
    \centering
    \caption{Effect of Job Displacement on Earnings including Additional Covariates}
    \label{fig:ife_covs}
    \begin{subfigure}[b]{.49\textwidth}
    \includegraphics[width=\textwidth]{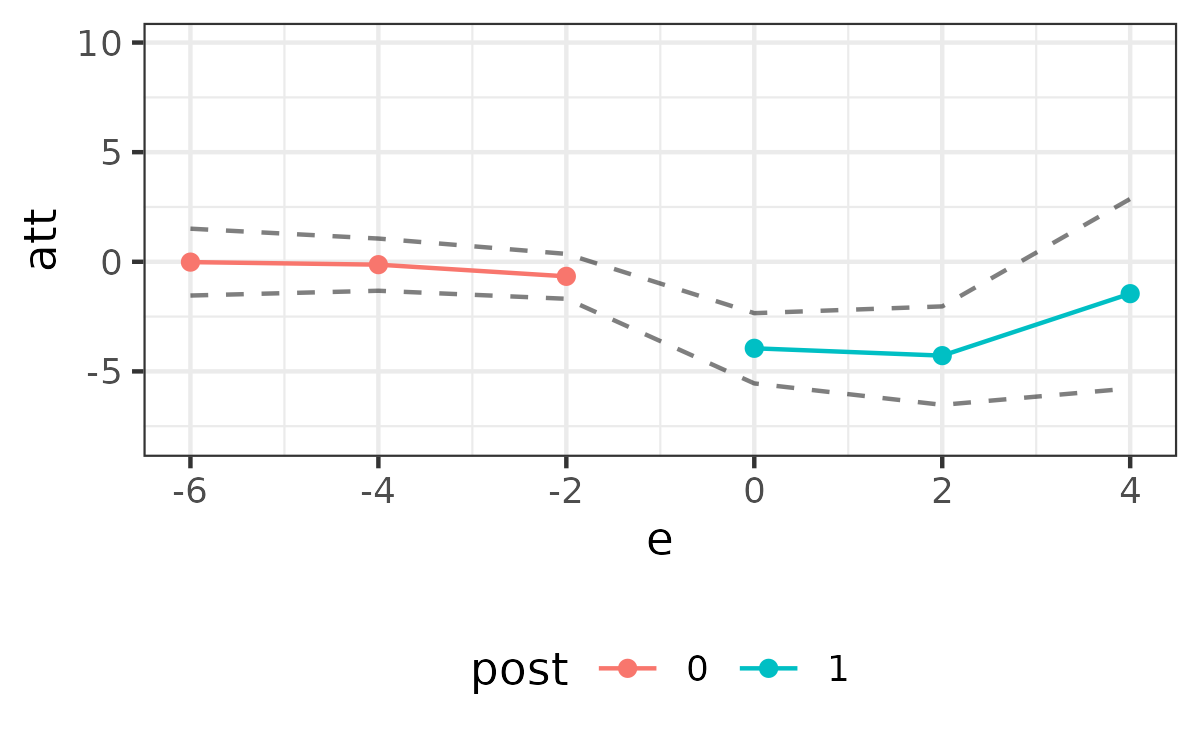}
    \caption{IFE, Time invariant AFQT}
    \end{subfigure}
    \begin{subfigure}[b]{.49\textwidth}
    \includegraphics[width=\textwidth]{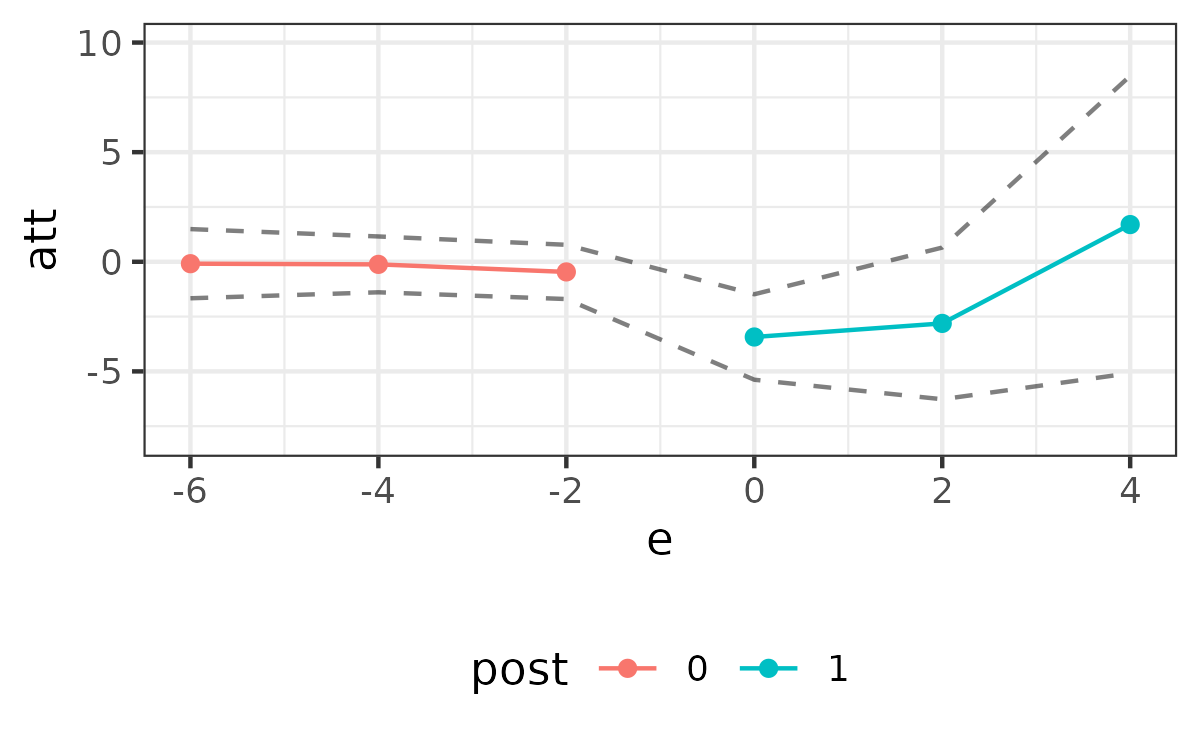}
    \caption{IFE, Time invariant education}
    \end{subfigure}
    \begin{subfigure}[b]{.49\textwidth}
    \includegraphics[width=\textwidth]{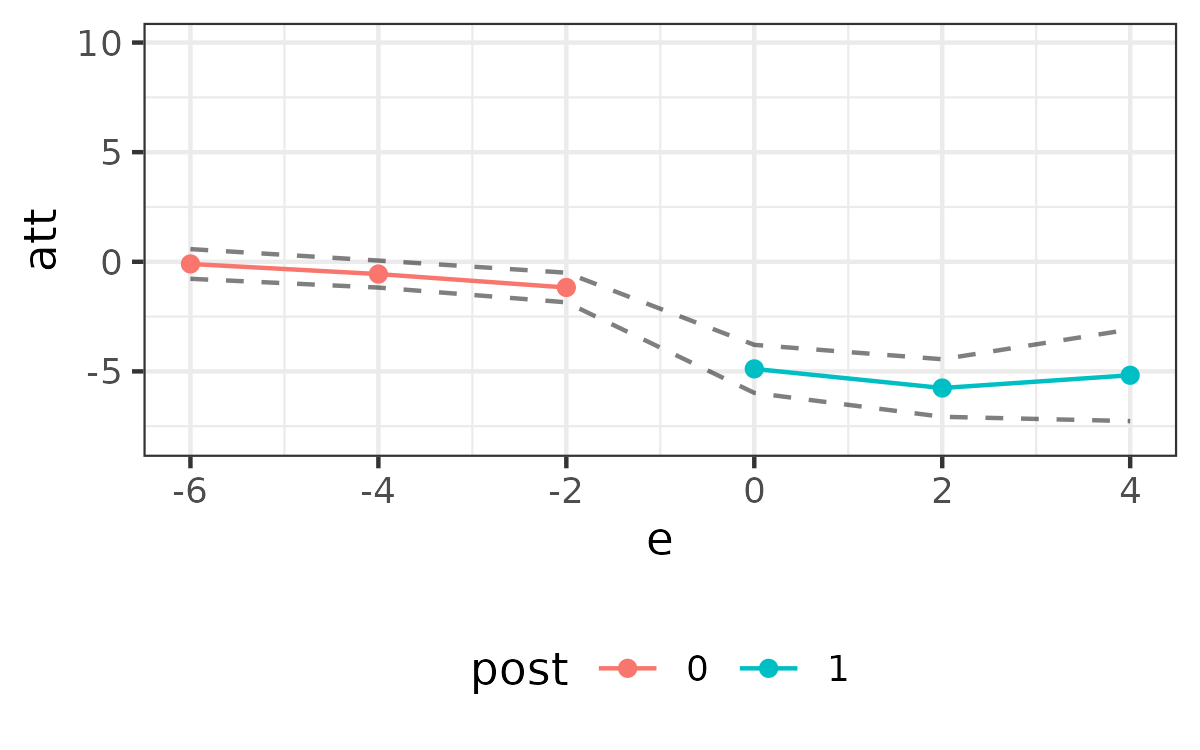}
    \caption{DID}
    \end{subfigure}
    \begin{subfigure}[b]{.49\textwidth}
    \includegraphics[width=\textwidth]{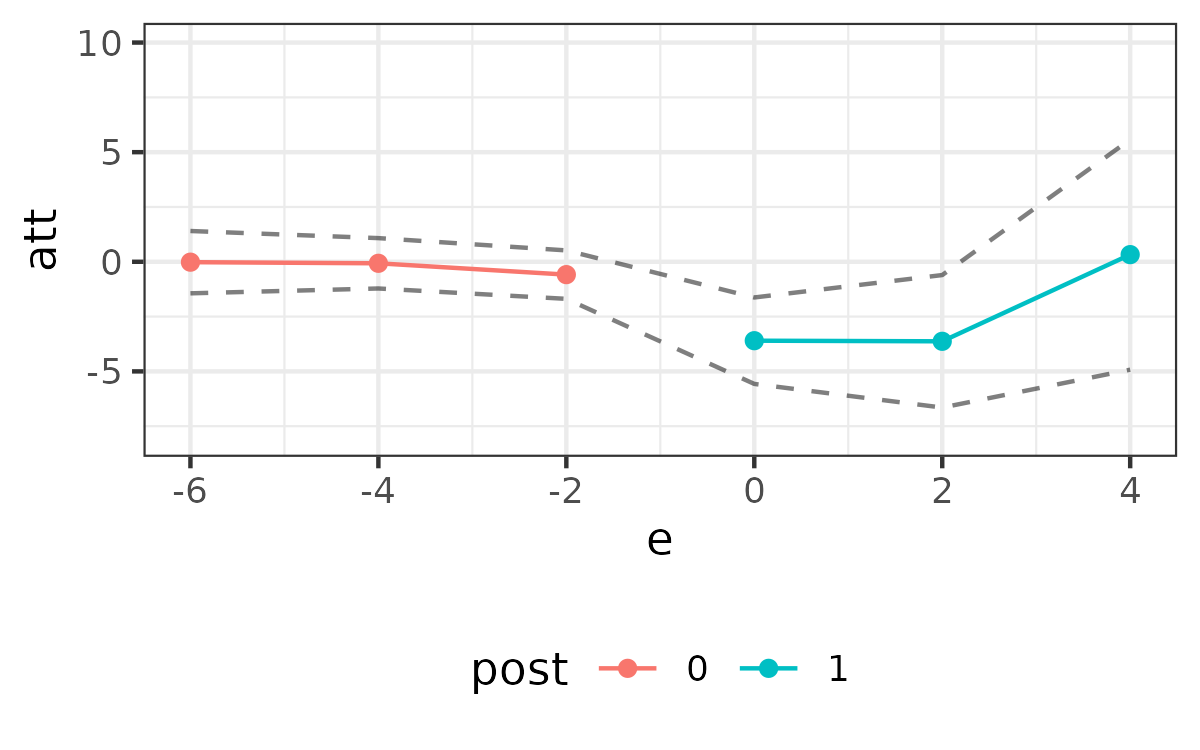}
    \caption{Linear Trends}
    \end{subfigure}
    
    \begin{justify}
    { \footnotesize \textit{Notes:} The figure contains estimates of the effect of job displacement on yearly earnings (which are in thousands of dollars) by length of exposure to job displacement.  Negative values of $e$ correspond to the number of years before job displacement occurred, $e=0$ in the year that job displacement occurred, and positive values of $e$ indicate how long ago an individual was displaced from their job.  Panels (a) and (b) contain estimates from the interactive fixed effects approach proposed in the current paper using AFQT (in panel (a)) or educational attainment (in panel (b)) as the covariates with time invariant effects on untreated potential outcomes.  Panels (c) and (d) contain analogous estimates but using difference in differences and linear trends models, respectively.  All panels additionally include educational attainment (except panel (b)), race, and gender as covariates that can have time varying effects on untreated potential outcomes.  Each panel provides 90\% confidence intervals computed using the multiplier bootstrap.
    }
    \end{justify}
\end{figure}

\begin{figure}[t]
    \centering
    \caption{Effect of Job Displacement on Earnings with Multiple Interactive Fixed Effects}
    \label{fig:multiple-ifes}
    \begin{subfigure}[b]{.49\textwidth}
    \includegraphics[width=\textwidth]{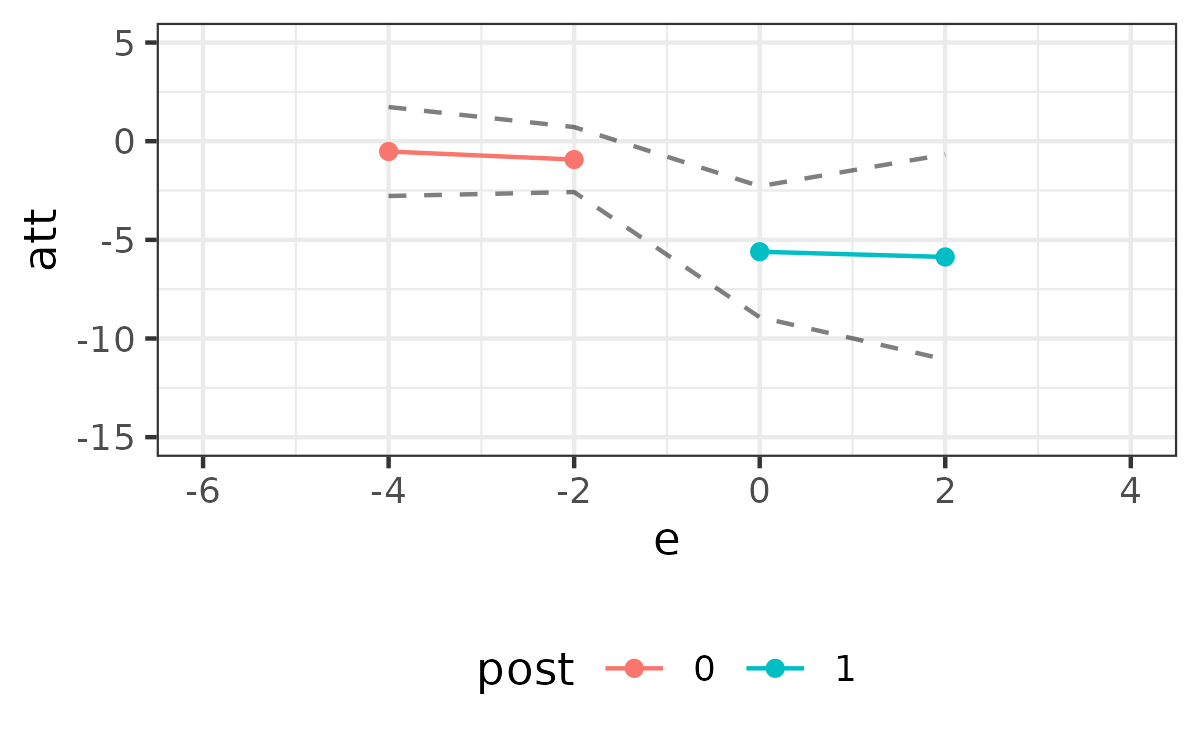}
    \caption{Two IFEs, no covariates}
    \end{subfigure}
    \begin{subfigure}[b]{.49\textwidth}
    \includegraphics[width=\textwidth]{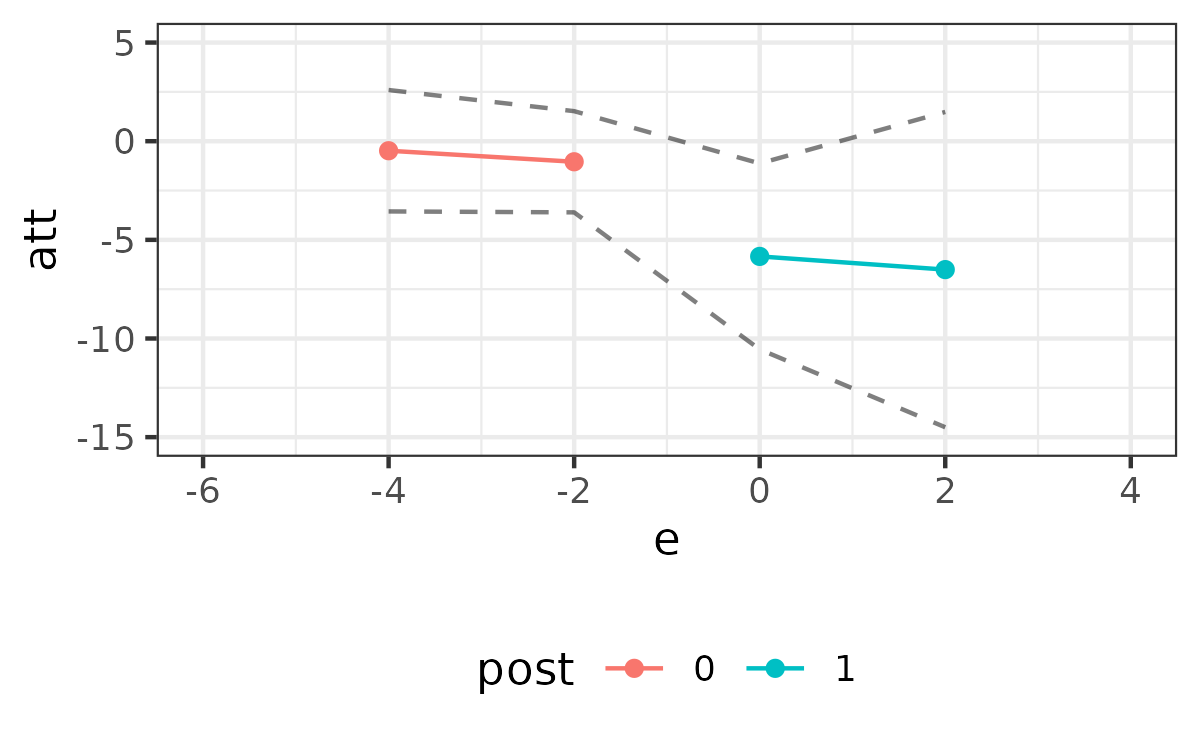}
    \caption{Two IFEs, covariates}
    \end{subfigure}
    \begin{justify}
    { \footnotesize \textit{Notes:} The figure contains estimates of the effect of job displacement on yearly earnings (which are in thousands of dollars) by length of exposure to job displacement.  Negative values of $e$ correspond to the number of years before job displacement occurred, $e=0$ in the year that job displacement occurred, and positive values of $e$ indicate how long ago an individual was displaced from their job.  Panel (a) contains estimates from the interactive fixed effects approach proposed in the current paper with two interactive fixed effects and using AFQT and educational attainment as covariates with time invariant effects on untreated potential outcomes.  Panel (b) contains analogous estimates but additionally includes race and gender as covariates that can have time varying effects on untreated potential outcomes.  Each panel provides 90\% confidence intervals computed using the multiplier bootstrap.
    }
    \end{justify}
\end{figure}

\end{document}